\documentclass[11pt]{article}

\usepackage{amsmath}
\usepackage{graphicx}
\usepackage{amsfonts}
\usepackage{amssymb}
\usepackage{epsfig}
\usepackage{color}
\usepackage{psfrag}

\begin{document}
\setcounter{page}{0} \topmargin 0pt
\renewcommand{\thefootnote}{\arabic{footnote}}
\newpage
\setcounter{page}{0}

\begin{titlepage}

\vspace{0.5cm}
\begin{center}
{\Large {\bf Kink scaling functions in 2D}}\\
{\Large {\bf non--integrable quantum field theories}}


\vspace{2cm}
{\large G. Mussardo$^{1,2}$, V. Riva$^{3}$, G. Sotkov$^{4,*}$ and G. Delfino$^{1,2}$} \\
\vspace{0.5cm} {\em $^{1}$International School for Advanced Studies}\\
{\em Via Beirut 1, 34100 Trieste, Italy} \\
\vspace{0.3cm} {\em $^{2}$Istituto Nazionale di Fisica Nucleare, Sezione di Trieste}\\
\vspace{0.3cm} {\em $^3$ Rudolf Peierls Centre for Theoretical
Physics, University of Oxford}\\ {\em 1 Keble Road, Oxford, OX1
3NP, UK}\\{\em and Wolfson College, Oxford}
\\\vspace{0.3cm} {\em $^4$ Departamento de Fisica, Universidade
Federal do Espirito Santo}\\
{\em 29060-900 Vitoria, Espirito Santo, Brazil}
\end{center}
\vspace{1cm}

\begin{abstract}
We determine the semiclassical energy levels for the $\phi^4$ field theory in
the broken symmetry phase on a 2D cylindrical geometry with
antiperiodic boundary conditions by quantizing the
appropriate finite--volume kink solutions. The analytic form of
the kink scaling functions for arbitrary size of the system allows
us to describe the flow between the twisted sector of $c=1$ CFT in
the UV region and the massive particles in the IR limit.
Kink-creating operators are shown to correspond in the UV limit to
disorder fields of the $c=1$ CFT. The problem of the
finite--volume spectrum for generic 2D Landau--Ginzburg models is
also discussed.

\vspace{2cm}  \hrulefill

E-mail addresses: mussardo@sissa.it,\,\,
v.riva1@physics.ox.ac.uk,\,\, sotkov@inrne.bas.bg \,\,
delfino@sissa.it

\vspace{0.5cm}

$^*$ On leave of absence - Institute for Nuclear Research and
Nuclear Energy, Bulgarian Academy of Sciences, Tsarigradsko
Chaussee 72, BG-1784, Sofia, Bulgaria.

\end{abstract}

\end{titlepage}

\newpage

\section{Introduction}\setcounter{equation}{0}

The universal thermodynamical properties of statistical systems
with multicritical behavior are described, in mean--field
approximation, by appropriate Landau--Ginzburg (LG) field
theories:
\begin{equation}\label{LGpot}
V_l(\phi)\,=\,\sum\limits_{k=1}^{l}\,\lambda_k\phi^{2k-2}\;\;,\qquad\qquad
l=3,4,...
\end{equation}
Structural (commensurate--incommensurate) phase transitions
\cite{transferintegral}, interface phenomena in ordered and
disordered media \cite{jasnov} and phase structure of
ferromagnetic systems (see for instance \cite{zinnjustin}) provide
few examples for the applications of the simplest $\phi^4$ and
$\phi^6$ LG models to statistical mechanics and condensed matter
physics. In two dimensions, the LG potentials (\ref{LGpot}) appear
also in the description of the relevant perturbations of Virasoro
minimal models of conformal field theory \cite{LG}, as well as of
the renormalization group flows between them.

The physical quantities associated with a field theory
--- partition function, energy spectrum, correlation functions, etc.
--- strongly depend on the geometry of the considered problem
(cylindrical, strip, plane, etc.), on the boundary conditions
chosen (periodic, Dirichlet, etc.) and on the range of the values
of the couplings $\lambda_k$. For several integrable quantum field
theories in 2D, the above quantities have been exactly computed in
finite volume with the so-called Thermodynamics Bethe Ansatz
method \cite{AlTBA} or Destri--deVega equations \cite{DdV}. These
techniques, however, require the integrability of the model, and
cannot be applied to the LG theories (\ref{LGpot}), due to their
non--integrable nature. In this case, the analysis of the
finite--size effects is based on approximative methods as
perturbative renormalization group (see \cite{jasnov,zinnjustin}
and references therein), transfer integral techniques
\cite{transferintegral} and numerical methods.

The low temperature (broken symmetry) phase of these models
exhibits, however, specific features --- multiple degenerate
vacua, non--trivial topological sectors and non--perturbative kink
solutions (domain walls) --- which require certain improvements of
the standard perturbative methods. The non--perturbative
semiclassical expansion \cite{DHN} is known to be an effective
method for the quantization of the kink solutions in an infinite
volume, independently of the integrability of the model. Its
recent extension to finite geometries \cite{SGscaling,SGstrip}
allowed us to derive analytic expressions for the scaling
functions of the Sine--Gordon model defined on a cylinder with
quasi--periodic b.c. (i.e. in the one--kink sector) and on a strip
with Dirichlet b.c.'s. It is then natural to address the problem
of the finite--size effects in 2D LG models within the context of
the semiclassical quantization of kinks in finite volume.

The present paper is devoted to the derivation of the scaling
functions of the 2D $\phi^4$ theory on a cylindrical geometry with
\textit{antiperiodic} b.c. $\phi(x+R)=-\phi(x)$, which for this
model corresponds to consider a single kink on the cylinder. This
continues our analysis of finite-size effects in the $\phi^4$
model, which begun in \cite{finvolff} with the derivation of the
finite--volume form factors and spectral functions for the same
kind of geometry.

From the mathematical point of view, the derivation of the scaling
functions for the $\phi^4$ theory on the \textit{twisted} cylinder
is analogous to the one performed in \cite{SGscaling} for the
Sine--Gordon model on a cylinder with \textit{quasi-periodic}
b.c.. This is due to the fact that the finite volume kinks are
expressed in both cases in terms of a Jacobi elliptic function,
and the computation of the corresponding energy levels is
therefore based on the solution of the so--called Lam\'e equation.
Besides a minor technical difference (the equation appears now in
a more complicated form, the so--called $N=2$ Lam\'e form), an
important new feature emerges in the antiperiodic case: the
oscillating background cannot be defined for any value of the size
of the system, so that the complete description of the problem is
achieved in this case by also including a constant background below a specific
value of the size.

Our main result, presented in Sect.\,\ref{sectphi4}, consists in
the analytic expression of the kink scaling functions (for
arbitrary value of the size of the system $R$), which describes the
flow between the twisted sector of $c=1$ CFT in the UV region and
the massive particles in the $Q=\pm 1$ topological sectors of the
broken $\phi^4$ theory in the infrared (IR) limit. This Section also
includes a comparison between the large--$R$ corrections to the
kink masses, as obtained from the IR asymptotic behaviour of the
scaling functions, and the values expected from the
infinite--volume scattering data through Luscher's theory
\cite{luscher}.

A detailed study of the UV regime is left to
Sect.\,\ref{sectkinkop}. Here we analyse the  properties of the
$c=1$ CFT fields that play the role of creating operators for the
$\phi^4$ kinks, as well as of the kinks of generic LG models. It
turns out that for $Z_2$--invariant polynomial potentials (in
their broken phase) the disorder field $\mu$ of dimension $1/8$
(and its descendants) from the twisted sector of the $c=1$ CFT are
the only operators local with respect to the potential and
carrying topological ($Z_2$) charges. Therefore they  must
describe the UV limit of the LG--kinks.

Sect.\,\ref{sectkinkop} actually begins with the more familiar
discussion of soliton--creating operators for the Sine-Gordon
model in the winding (i.e. quasiperiodic) sector. Due to the
compactification of the field, indeed, this theory admits more
types of b.c., including the antiperiodic ones. We have then
devoted Sect.\,\ref{sectSG} to the analysis of this interestingly
rich model, which displays two types of non-trivial classical
solutions in finite volume, respecting two different b.c.'s
(quasiperiodic and antiperiodic). Their UV limits are described,
respectively, by the standard soliton--creating operators from the
winding sector of $c=1$ CFT and by the disorder field in its
twisted sector, i.e. that one which creates the $Z_2$ charged kinks.
The two corresponding types of scaling functions are given explicitly, and
their difference is observed at any finite volume, except for
their identical IR limits. It is therefore clear that passing from
periodic to $Z_2$--symmetric polynomial potentials only the
kink--type (antiperiodic) solution survives, which explain why the
finite volume kink--type solutions of SG and $\phi^4$ models (as
well as their UV limits) share many common properties.

The explicit analytic form obtained in the present paper for the scaling functions
of the $\phi^4$ model (and in previous works
\cite{SGscaling,SGstrip} for the Sine--Gordon model) is
intrinsically related to the fact that the stability equations to
be solved are of Lam\'e type, and the corresponding solutions are
well known. As we shall show in Sect.\,\ref{sectphiN}, similar
construction for $\phi^6$ and higher ($l\geq 5$) LG models leads
again to Schr\"odinger--like equations for periodic potentials,
but it turns out that these are more complicated generalizations
of the Lam\'e equation. The derivation of the finite--volume energy spectrum of
these models thus depends on the further progress that will be
achieved in the future on their analytical or numerical solutions.

\section{Semiclassical quantization of the broken $\phi^4$ theory in finite
volume}\label{sectphi4} \setcounter{equation}{0}

The standard perturbative methods of QFT's in $D$-dimensions
(including the $D=2$ case we are interested in) are known to be
inefficient for the description of the quantum effects in the
topologically non--trivial sectors of an important class of
theories with non--linear interactions and multiple
degenerate vacua. As a rule, such theories admit finite--energy
non--perturbative classical solutions (kinks, vortices, monopoles
etc.) carrying topological charges. The quantization of these
solutions (both static and time--dependent) requires
non--perturbative techniques, developed by Dashen, Hasslacher and
Neveu (DHN) in \cite{DHN} for theories in infinite volume. The DHN
method consists, for static backgrounds, in splitting the field
$\phi(x,t)$ in terms of the classical solution and its quantum
fluctuations, i.e.
\begin{equation*}
\phi(x,t) \,=\,\phi_{cl}(x) + \eta(x,t) \,\,\, \,\,\, , \,\,\,
\,\,\, \eta(x,t) \,=\,\sum_{k} e^{i \omega_k t} \,\eta_{k}(x)
\,\,\,,
\end{equation*}
and in further expanding the Lagrangian of the theory in powers of
$\eta$, keeping only the quadratic terms. As a result of this
procedure, $\eta_{k}(x)$ satisfies the so called \lq\lq stability
equation"
\begin{equation}\label{stability}
\left[-\frac{d^2}{d x^2} + V''(\phi_{cl}) \right] \, \eta_{k}(x)
\,=\, \omega_k^2 \,\eta_k(x) \,\,\,,
\end{equation}
together with certain boundary conditions. The semiclassical
energy levels in each sector are then built in terms of the energy
of the corresponding classical solution and the eigenvalues
$\omega_i$ of the Schr\"odinger--like equation (\ref{stability}),
i.e.
\begin{equation}
E_{\{n_i\}} \,=\,{\cal E}_{cl} + \hbar \,\sum_{k}\left(n_k +
\frac{1}{2}\right) \, \omega_k + O(\hbar^2) \,\,\,, \label{tower}
\end{equation}
where $n_k$ are non--negative integers. In particular the ground
state energy in each sector is obtained by choosing all $n_k = 0$
and it is therefore given by\footnote{From now on we will fix
$\hbar=1$, since the semiclassical expansion in $\hbar$ is
equivalent to the expansion in the interaction coupling
$\lambda$.}
\begin{equation}
E_{0} \,=\,{\cal E}_{cl} + \frac{\hbar}{2} \,\sum_{k} \omega_k +
O(\hbar^2) \,\,\,. \label{e0}
\end{equation}

In our recent papers \cite{SGscaling,SGstrip}, we have extended
this technique to the study of soliton quantization in the
Sine--Gordon model on the cylinder (with periodic b.c.) and on a
strip with Dirichlet b.c.. This Section is devoted to the
quantization of the kinks of the $\phi^4$ theory in the
$\mathbb{Z}_2$ broken symmetry phase, defined by the Lagrangian
\begin{equation}\label{phi4pot}
{\cal L}\,=\,\frac{1}{2}\left(\partial_\mu
\phi\right)\left(\partial^\mu
\phi\right)-V(\phi)\;,\qquad\text{with}\qquad
V(\phi)\,=\,\frac{\lambda}{4}\left(\phi^2-\frac{m^2}{\lambda}\right)^2\;,
\end{equation}
on a cylinder with the antiperiodic b.c.'s
\begin{equation}\label{antiperbc}
\phi(x+R)=-\phi(x)\;,
\end{equation}
imposed. In order to fix the ideas and the notations, we first shortly
review the DHN method for the quantization of $\phi^4$--kinks in
infinite volume.

\subsection{Infinite volume kinks}

The static solutions of the equation of motion associated to the
potential (\ref{phi4pot}) can be obtained by integrating the
following first order equation
\begin{equation}\label{firstorderphi4}
\frac{1}{2}\left(\frac{\partial \bar{\phi}_{cl}}{\partial
\bar{x}}\right)^{2}=\frac{1}{4}\left(\bar{\phi}_{cl}^{2}-\bar{\phi}_{0}^{2}\right)
\left(\bar{\phi}_{cl}^{2}-2+\bar{\phi}_{0}^{2} \right)\;,
\end{equation}
where we have rescaled the variables as
\begin{equation}\label{scaledvarphi4}
\bar{\phi}=\frac{\sqrt{\lambda}}{m}\,\phi\;,\qquad \qquad\bar{x}=m
x\;,
\end{equation}
and $\phi_{0}$ is an arbitrary constant defined by
$V(\phi_{0})=-A$, i.e.
\begin{equation*}
\frac{1}{2}\left(\frac{\partial \phi_{cl}}{\partial x}\right)^{2}
\,=\, V(\phi_{cl}) + A \,\,\,. \label{firstorder}
\end{equation*}

\begin{figure}[ht]
\begin{tabular}{p{8cm}p{7cm}}
\psfrag{phi}{$\bar{\phi}$}\psfrag{V}{$V(\bar{\phi})$}\psfrag{b}{\small$1$}
\psfrag{a}{\small$\hspace{-0.3cm}-1$}
\psfig{figure=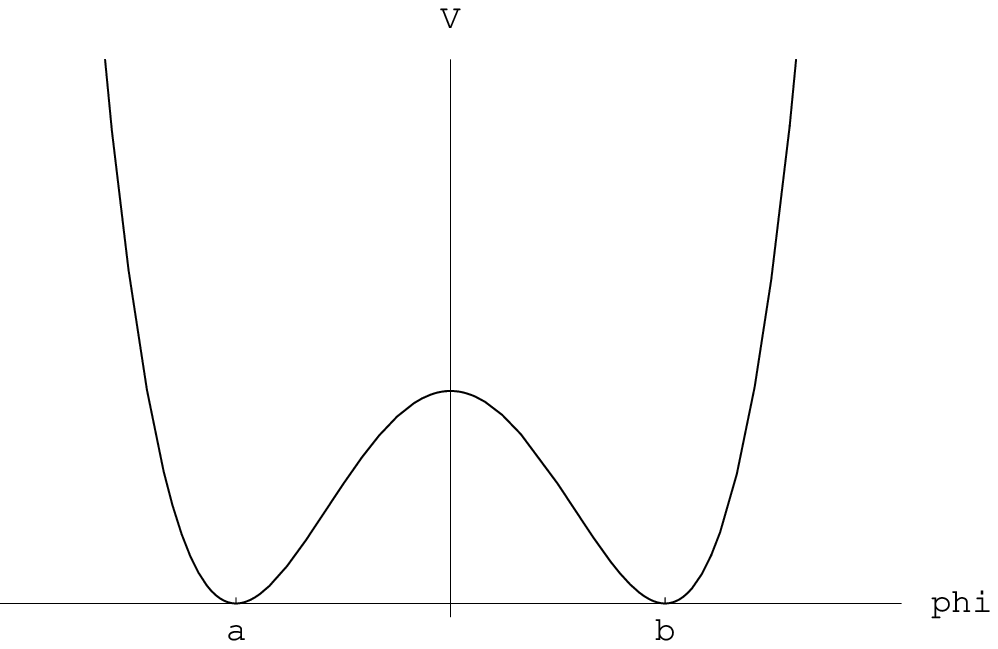,height=4cm,width=6.5cm} &
\psfrag{x}{$\bar{x}$}\psfrag{phi}{$\bar{\phi}_{cl}(x)$}\psfrag{c}{\small$\hspace{-0.2cm}1$}
\psfrag{d}{$\small\hspace{0.3cm}-1$}
\psfig{figure=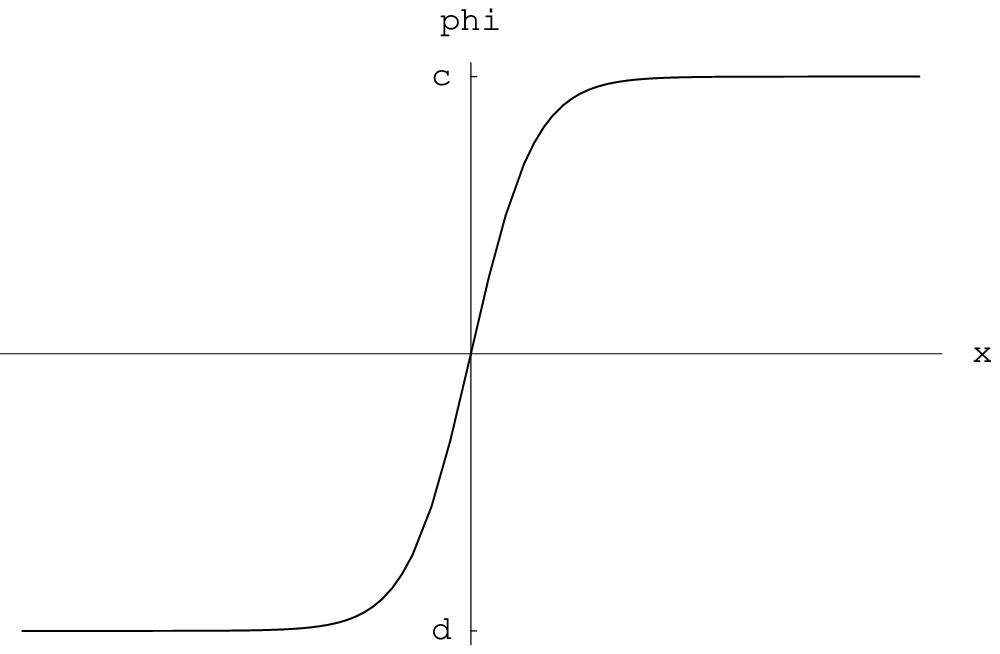,height=4cm,width=5.5cm}
\end{tabular}
\caption{Potential (\ref{phi4pot}) and infinite--volume kink
(\ref{phi4kinkinfvol}) with $x_0=0$.}\label{figphi4infvol}
\end{figure}

In infinite volume we have to impose as b.c. that the classical
field reaches the minima of the potential at $x\to\pm\infty$, i.e.
$\bar{\phi}_{cl}(\pm\infty)=\pm 1$. This corresponds to choosing
the value $\bar{\phi}_0=1$ for the arbitrary constant in
(\ref{firstorderphi4}), and, as a consequence, we find the
well--known kink solution
\begin{equation}\label{phi4kinkinfvol}
\bar{\phi}_{cl}(x)=\tanh\left(\frac{\bar{x}-\bar{x}_0
}{\sqrt{2}}\right)\;,
\end{equation}
shown in Fig.\,\ref{figphi4infvol}, which has classical energy
${\cal E}_{cl}=\frac{2\sqrt{2}}{3}\frac{m^{3}}{\lambda}$.

\vspace{0.5cm}

The stability equation (\ref{stability}) around this background
can be cast in the hypergeometric form in the variable
$z=\frac{1}{2}(1+\tanh \frac{\bar{x}}{\sqrt{2}})$, and the
solution is expressed in terms of the hypergeometric function
$F(\alpha,\beta,\gamma;\,z)$ as
\begin{equation*}
\eta(x)\,=\,z^{\sqrt{1-\frac{\omega^{2}}{2
m^{2}}}}(1-z)^{-\sqrt{1-\frac{\omega^{2}}{2 m^{2}}}}\,
F\left(3,-2,1+2\sqrt{1-\frac{\omega^{2}}{2 m^{2}}};\;z\right)\;.
\end{equation*}
The corresponding spectrum is given by the two discrete
eigenvalues
\begin{equation}\label{omega0phi4}
\omega_{0}^{2}=0\;,\qquad\text{with}\qquad
\eta_0(x)=\frac{1}{\cosh^2\frac{\bar{x}}{\sqrt{2}}}\;,
\end{equation}
and
\begin{equation}\label{omega1phi4}
\omega_{1}^{2}=\frac{3}{2}\,m^{2}\;,\qquad\text{with}\qquad
\eta_1(x)=\frac{\sinh\frac{\bar{x}}{\sqrt{2}}}{\cosh^2\frac{\bar{x}}{\sqrt{2}}}\;,
\end{equation}
plus the continuous part, labelled by $q\in\mathbb{R}$,
\begin{equation}\label{omegaqphi4}
\omega_{q}^{2}=m^{2}\left(2+\frac{1}{2}\,q^{2}\right)\;,\qquad\text{with}\qquad
\eta_q(x)=e^{i q
\bar{x}/\sqrt{2}}\left(3\tanh^2\frac{\bar{x}}{\sqrt{2}}-1-q^2-3 i
q \tanh\frac{\bar{x}}{\sqrt{2}}\right)\;.
\end{equation}
The presence of the zero mode $\omega_0$ is due to the arbitrary
position of the center of mass $x_0$ in (\ref{phi4kinkinfvol}),
while $\omega_1$ and $\omega_q$ represent, respectively, an
internal excitation of the kink particle and the scattering of the
kink with mesons\footnote{The mesons represent the excitations
over the vacua, i.e. the constant backgrounds
$\phi_{\pm}=\pm\frac{m}{\sqrt{\lambda}}$, therefore their square
mass is given by $V''(\phi_{\pm})=2\,m^2$.} of mass $\sqrt{2}\,m$
and momentum $m q/\sqrt{2}$.

The semiclassical correction to the kink mass can be now computed
as the difference between the ground state energy in the kink
sector and the one of the vacuum sector, plus a mass counterterm
due to normal ordering:
\begin{equation}\label{grstateinfvol}
M={\cal
E}_{cl}+\frac{1}{2}\,m\sqrt{\frac{3}{2}}+\frac{1}{2}\sum_{n}\left[m\sqrt{2+\frac{1}{2}\,q_{n}^{2}}-\sqrt{k_{n}^{2}
+2m^{2}}\right]-\frac{1}{2}\,\delta
m^{2}\int\limits_{-\infty}^{\infty}dx\left[\phi^{2}_{cl}(x)-\frac{m^{2}}{\lambda}\right]\;,
\end{equation}
with
\begin{equation}
\delta m^{2}=\frac{3\lambda}{4\pi}\int\limits_{-\infty}^{\infty}
\frac{d k}{\sqrt{k^{2}+2m^{2}}}\;.
\end{equation}
The discrete values $q_{n}$ and $k_{n}$ are obtained by putting the
system in a big finite volume of size $R$ with periodic boundary
conditions:
\begin{equation}
2 n\pi= k_{n}R = q_{n}\frac{m R}{\sqrt{2}}+\delta(q_{n})\;,
\end{equation}
where the phase shift $\delta(q)$ is extracted from $\eta_q(x)$ in
(\ref{omegaqphi4}) as
\begin{equation}
\eta_q
(x)\;{\mathrel{\mathop{\kern0pt\longrightarrow}\limits_{x\to
\pm\infty
}}}\;e^{i\left[q\,\frac{mx}{\sqrt{2}}\pm\frac{1}{2}\,\delta(q)\right]}\;,\qquad\qquad
\delta(q)=-2\arctan\left(\frac{3 q}{2-q^2}\right)\;.
\end{equation}
Sending $R\rightarrow\infty$ and computing the integrals one
finally has
\begin{equation}\label{kinkmass}
M\,=\,\frac{2\sqrt{2}}{3}\,\frac{m^{3}}{\lambda}+m\left(\frac{1}{6}\,\sqrt{\frac{3}{2}}-\frac{3}{\pi\sqrt{2}}\right)\;.
\end{equation}
Notice that, from the knowledge of this quantity, one can extract a rough
estimate of the value of couplings at which the broken $\phi^4$
theory actually describes the Ising model. It is well known, in
fact, that perturbing the conformal gaussian theory ${\cal
L}_{G}\,=\,\frac{1}{2}\left(\partial_\mu
\phi\right)\left(\partial^\mu \phi\right)$ with the potential
(\ref{phi4pot}) one can have different renormalization group
trajectories depending on the values of the dimensionless coupling
$\lambda/m^2$. The universality class of the Ising model is
described by the situation in which the infrared point is not a
massive theory but rather another conformal field theory, with central
charge $c=1/2$. Therefore, we can estimate semiclassically the
corresponding value of $\lambda/m^2$ by imposing the vanishing of
the mass (\ref{kinkmass}), which gives $\lambda/m^2\simeq 2$.
The large value of this quantity suggests, however, that the one--loop order
in the semiclassical expansion in $\lambda/m^2$ can hardly be able
to detect the Ising fixed point.

\subsection{Classical solutions in finite volume}\label{sectclasssol}

Before discussing the kink solution on the cylinder, it is worth
briefly recalling that the DHN method can be also applied to the
constant solutions describing the vacua in the periodic sector of
the theory. In particular, for the potential (\ref{phi4pot}) we
have
\begin{eqnarray}
&\phi_{cl}^{vac}(x)\equiv (\pm)\,\frac{m}{\sqrt{\lambda}}\;,\nonumber \\
&\omega_n^{vac}=\sqrt{2
m^2+\left(\frac{2n\pi}{R}\right)^{2}}\;,\qquad n=0,\pm 1,\pm
2...\;.
\end{eqnarray}
Therefore, according to (\ref{tower}), the smallest mass gap in
the system, i.e. the difference between the first excited state
and the ground state, is given by:
\begin{equation}
E_{1}(R)-E_{0}(R)=\omega_{0}^{vac}(R)\equiv\sqrt{2}\,m\;.
\end{equation}
This quantity, which is related to the inverse correlation length
$\xi^{-1}$ on a finite size \cite{zinnjustin,brezin,finitesize}, is the
one that has to be used\footnote{Up to inessential numerical constants which we
fix here to $1/\sqrt{2}$ for later convenience.} in the definition
of the scaling variable
\begin{equation}\label{scalvar}
r\,\equiv\,m\,R\;.
\end{equation}

\begin{figure}[ht]
\begin{tabular}{p{8cm}p{7cm}}
\psfrag{phi}{$\bar{\phi}$}\psfrag{V}{$V(\bar{\phi})$}\psfrag{b}{$$}
\psfrag{a}{$$}\psfrag{A}{\hspace{-0.2cm}\small$-A$}
\psfrag{phi0}{\footnotesize$\bar{\phi}_0$}\psfrag{fphi0}{\hspace{-0.5cm}\footnotesize$\sqrt{2-\bar{\phi}_0^2}$}
\psfig{figure=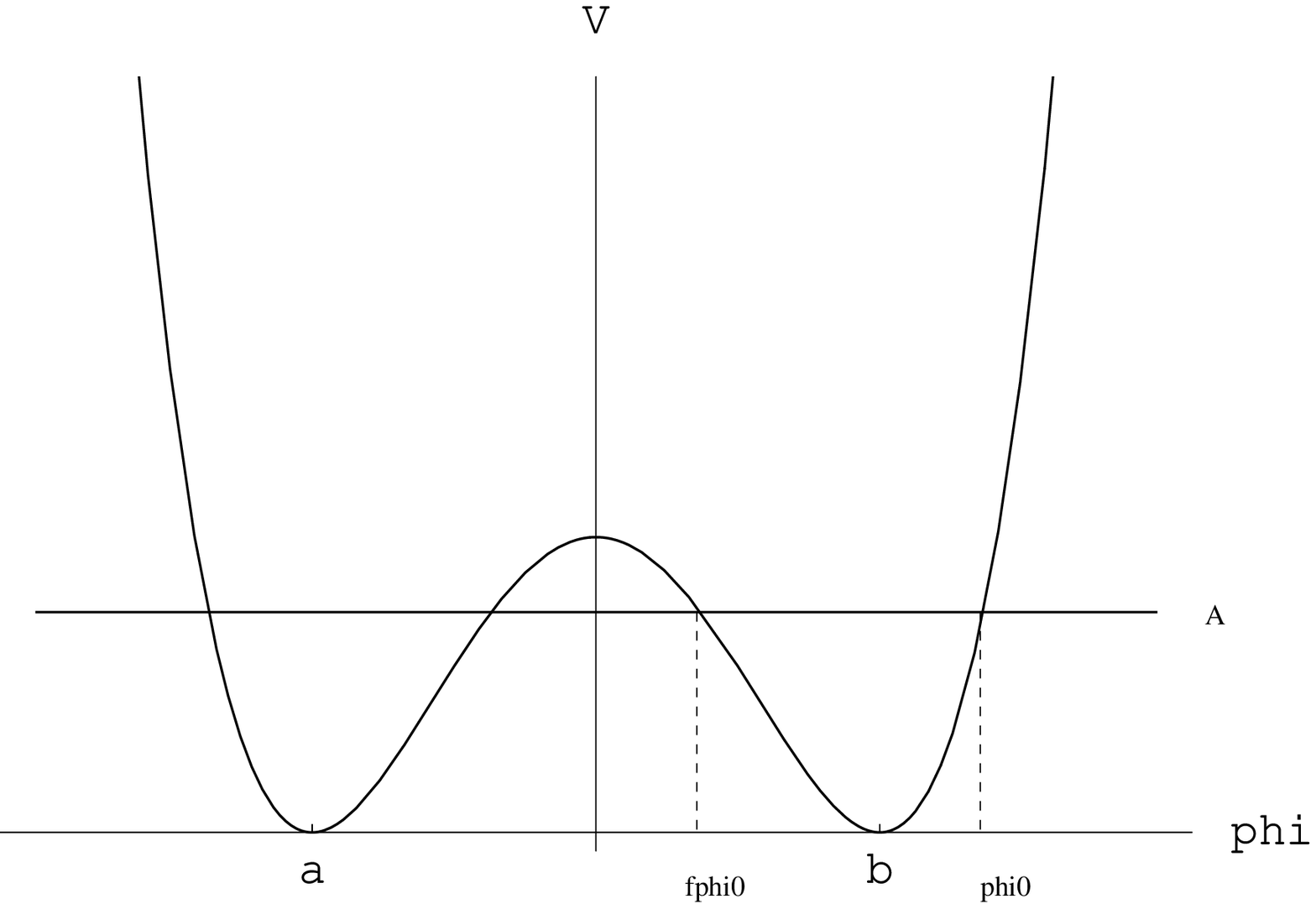,height=4cm,width=6.5cm} &
\psfrag{x}{$\frac{\bar{\phi}_{0}}{\sqrt{2}}\,
\,\bar{x}$}\psfrag{phicl(x)}{$\bar{\phi}_{cl}(\bar{x})$}
\psfrag{phi0}{\hspace{-1.8cm}\footnotesize$\sqrt{2-\bar{\phi}_{0}^{2}}$}
\psfrag{-phi0}{\footnotesize$-\sqrt{2-\bar{\phi}_{0}^{2}}$}
\psfrag{K(k^2)}{$\textbf{K}$}\psfrag{-K(k^2)}{$-\textbf{K}$}
\psfig{figure=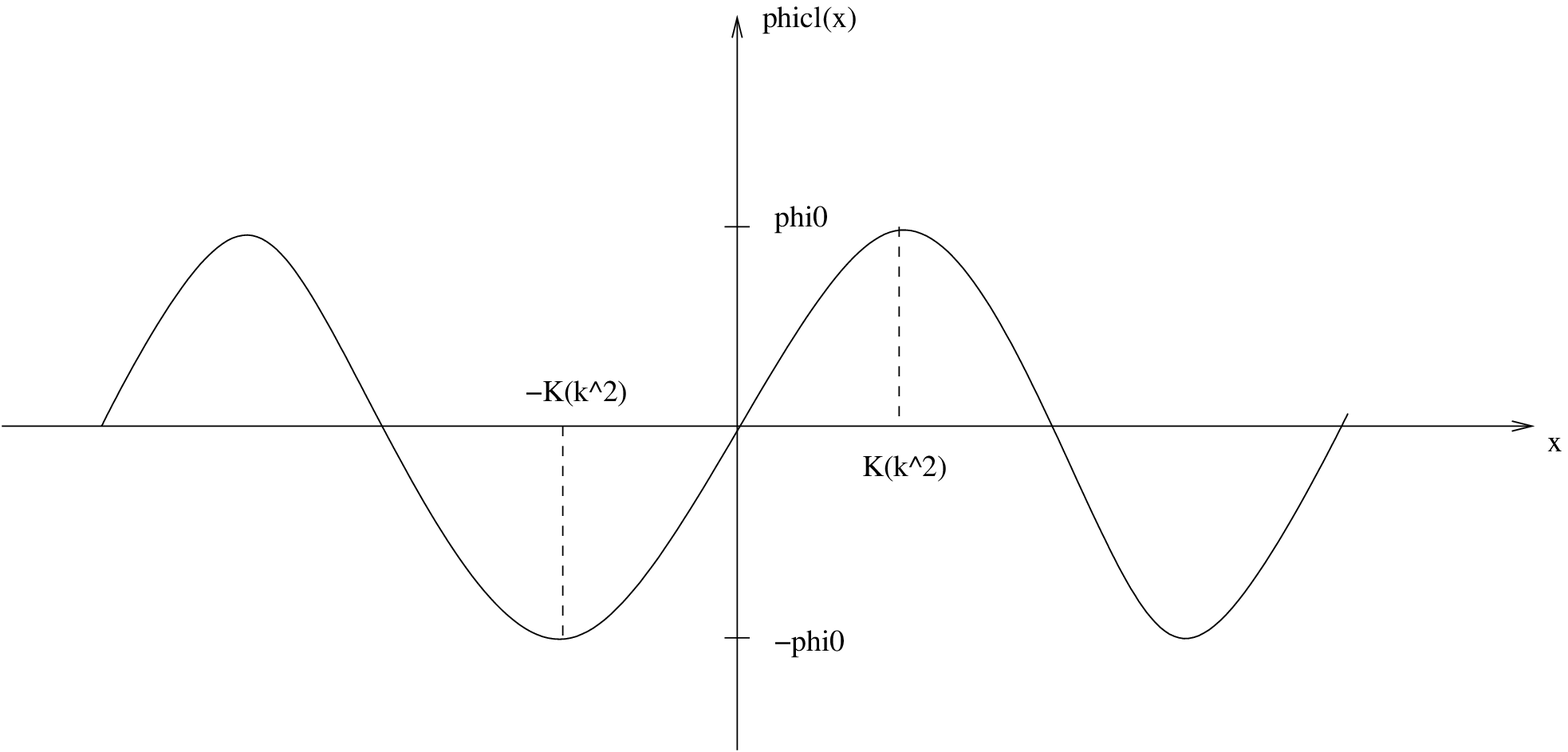,height=4cm,width=5.5cm}
\end{tabular}
\caption{Potential (\ref{phi4pot}) and finite--volume kink
(\ref{phi4kink}) with $x_0=0$.}\label{figphi4finvol}
\end{figure}

If we now want to describe a kink on a cylinder of circumference
$R$, we have to look for a solution of eq.\,(\ref{firstorderphi4})
satisfying the antiperiodic boundary conditions (\ref{antiperbc}).
This can be found for $1<\bar{\phi}_{0}<\sqrt{2}$, and it is
expressed as
\begin{equation}\label{phi4kink}
\bar{\phi}_{cl}(\bar{x})=\sqrt{2-\bar{\phi}_{0}^{2}}\;\;\textrm{sn}\left(\frac{\bar{\phi}_{0}}{\sqrt{2}}\,
\,(\bar{x}-\bar{x}_0)\,,\,k\right)\;,
\end{equation}
where $\text{sn}(u,k)$ is the Jacobi elliptic function with
modulus $k^{2}=\frac{2}{\bar{\phi}_{0}^{2}}-1$ and period
$4\textbf{K}(k^{2})$, where $\textbf{K}(k^{2})$ is the complete
elliptic integral of the first kind (see Appendix\,\ref{appell}
for the definitions and properties of elliptic integrals and
Jacobi elliptic functions). As shown in Fig.\,\ref{figphi4finvol},
the classical solution (\ref{phi4kink}) oscillates between the
values $-\sqrt{2-\bar{\phi}_{0}^{2}}$ and
$\sqrt{2-\bar{\phi}_{0}^{2}}$, and the boundary conditions
(\ref{antiperbc}) are satisfied by relating the elliptic modulus
to the size of the system as
\begin{equation}\label{sizephi4}
mR\,=\,\sqrt{1+k^2}\;\,2\,\textbf{K}(k^{2})\;.
\end{equation}

As expected, (\ref{phi4kink}) goes to the infinite--volume kink
(\ref{phi4kinkinfvol}) for $k\rightarrow 1$ ( i.e.
$\bar{\phi}_{0}\rightarrow 1$), which corresponds to the infrared
limit $mR\rightarrow\infty$. In the complementary limit
$k\rightarrow 0$ ( i.e. $\bar{\phi}_{0}\rightarrow \sqrt{2}$),
which corresponds to $mR \rightarrow \pi$, the kink
(\ref{phi4kink}) tends to the constant solution
\begin{equation}\label{phi4kinkUV}
\phi_{cl}(x)\equiv 0\;\;,
\end{equation}
which identically satisfies the antiperiodic b.c.
(\ref{antiperbc}) and can be used, therefore, as the background
field configuration in the interval $0 < m R < \pi$. The choice
of the background
\begin{equation}\label{phi4kinkcomplete}
\phi_{cl}(x)=\begin{cases}\sqrt{2-\bar{\phi}_{0}^{2}}\;\;\textrm{sn}\left(\frac{\bar{\phi}_{0}}{\sqrt{2}}\,
\,(\bar{x}-\bar{x}_0)\,,\,k\right)&\quad\text{for} \quad mR>\pi\\
0 &\quad\text{for} \quad mR<\pi\end{cases}\;
\end{equation}
will be fully motivated in the following, after the discussion of
the stability frequencies related to the classical solutions
(\ref{phi4kink}) and (\ref{phi4kinkUV}).

\begin{figure}[ht]
\begin{center}
\psfig{figure=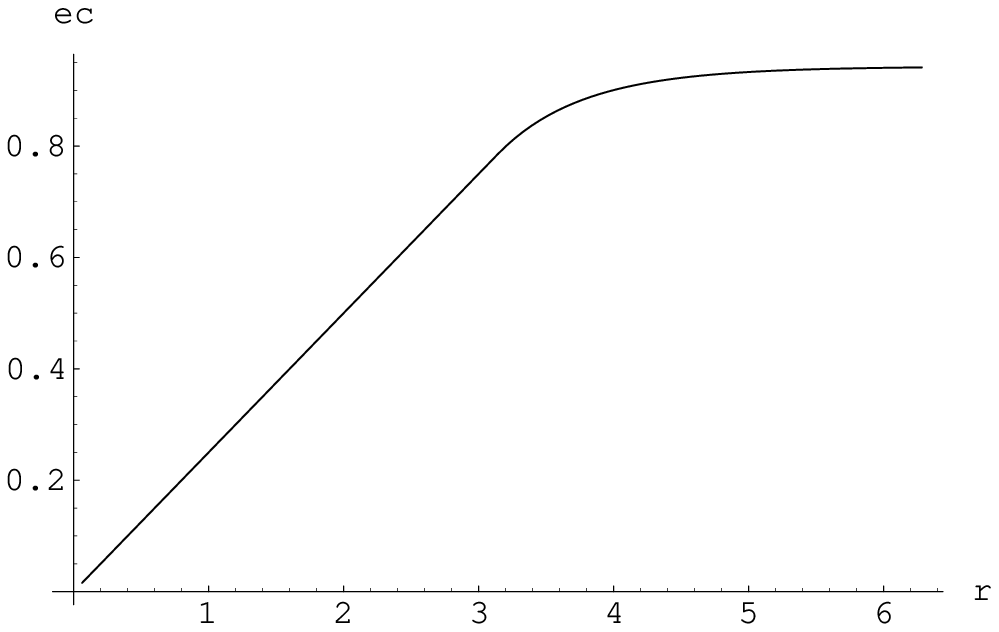,height=5cm,width=6.5cm} \caption{Classical
energy (\ref{phi4classen})}\label{figclassen}
\end{center}
\end{figure}

The classical energy of the kink (\ref{phi4kinkcomplete}) is given
by
\begin{equation}\label{phi4classen}
{\cal
E}_{cl}(R)=\begin{cases}\frac{m^{3}}{6\lambda}\,\frac{1}{\left(1+k^2\right)^{3/2}}
\left\{3 k^4 \textbf{K}(k^{2}) +2 k^2 \left[
\textbf{K}(k^{2})+4\textbf{E}(k^{2})\right]+8\textbf{E}(k^{2})-5\textbf{K}(k^{2})\right\}&\quad\text{for} \quad mR>\pi\\
\frac{m^{3}}{4\lambda}\, mR &\quad\text{for} \quad
mR<\pi\end{cases}\;,
\end{equation}
and it is plotted in Fig.\,\ref{figclassen}. From the analytic
knowledge of this quantity, we can immediately extract some
important scattering data of the non--integrable $\phi^4$ theory.
In fact, the leading term in the kink mass is given by the
classical energy, expressed for generic $R$ by
(\ref{phi4classen}). It is easy to see that for
$R\rightarrow\infty$ the energy indeed tends to the
infinite--volume limit ${\cal E}_{cl}(R)\rightarrow
\frac{2\sqrt{2}}{3}\frac{m^{3}}{\lambda}$. From its asymptotic
expansion for large $R$, we can also obtain the leading order of the
kink mass correction in finite volume, and compare it with
L\"{u}scher's theory \cite{luscher,kmluscher}. Taking into account
the $k\rightarrow 1$ ($k'\rightarrow 0$) expansions of
$\textbf{E}$ and $\textbf{K}$ (see Appendix\,\ref{appell}) and
noting from (\ref{sizephi4}) that
$$
e^{-\sqrt{2}m R}=\frac{1}{256}(k')^{4}+\cdots\;,
$$
we derive the following asymptotic expansion of ${\cal E}_{cl}$
for large $R$:
\begin{equation}\label{phi4classenIRexp}
{\cal E}_{cl}(R)={\cal
E}_{cl}(\infty)-8\sqrt{2}\,\,\frac{m^{3}}{\lambda} e^{-\sqrt{2}m
R}+O\left(e^{-2 \sqrt{2}m R}\right)\;.
\end{equation}
The counterpart of this leading--order behavior in L\"uscher's
theory is given by
\begin{equation}
M_k(R)-M_k(\infty)\,=\,-\,
m_{b}\,R_{k\,k\,b}\,e^{-m_{b}R}\;,\label{luscherR}
\end{equation}
where the index $k$ refers to the kink, and the index $b$ refers
to the elementary meson (with mass $m_b=\sqrt{2}\,m$), which can
be seen as a kink--antikink bound state with $S$--matrix residue
$R_{k\,k\,b}$. From the comparison between
(\ref{phi4classenIRexp}) and (\ref{luscherR}) we finally extract
the leading semiclassical expression for the residue of this
$3$--particle process
\begin{equation}\label{Rkkb}
R_{k\,k\,b}\,=\,8\,\frac{m^2}{\lambda}\;,
\end{equation}
and therefore the 3-particle coupling\footnote{Crossing symmetry
implies the equality $R_{k\bar{k}b}=R_{kkb}$.}
\begin{equation}\label{Gkkb}
\Gamma_{k\,\bar{k}\,b}=2\sqrt{2}\frac{m}{\sqrt{\lambda}}\;.
\end{equation}
This quantity is of particular interest, since the
non--integrability of the $\phi^4$ theory prevents the knowledge
of its exact $S$--matrix. In the different context of infinite
volume form factors, in \cite{finvolff} we proposed another way of
extracting this coupling, i.e. by looking at the residue of the
kink--antikink form factor in infinite volume, and the result obtained
in \cite{finvolff} is consistently equal to (\ref{Gkkb}).

\vspace{0.5cm}

\psfrag{ec}{$\frac{{\cal E}_{cl}}{m^3/\lambda}$}\psfrag{r}{$r$}

\subsection{Semiclassical scaling functions}\label{sectomegas}

\begin{figure}[h]
\hspace{3cm} \psfig{figure=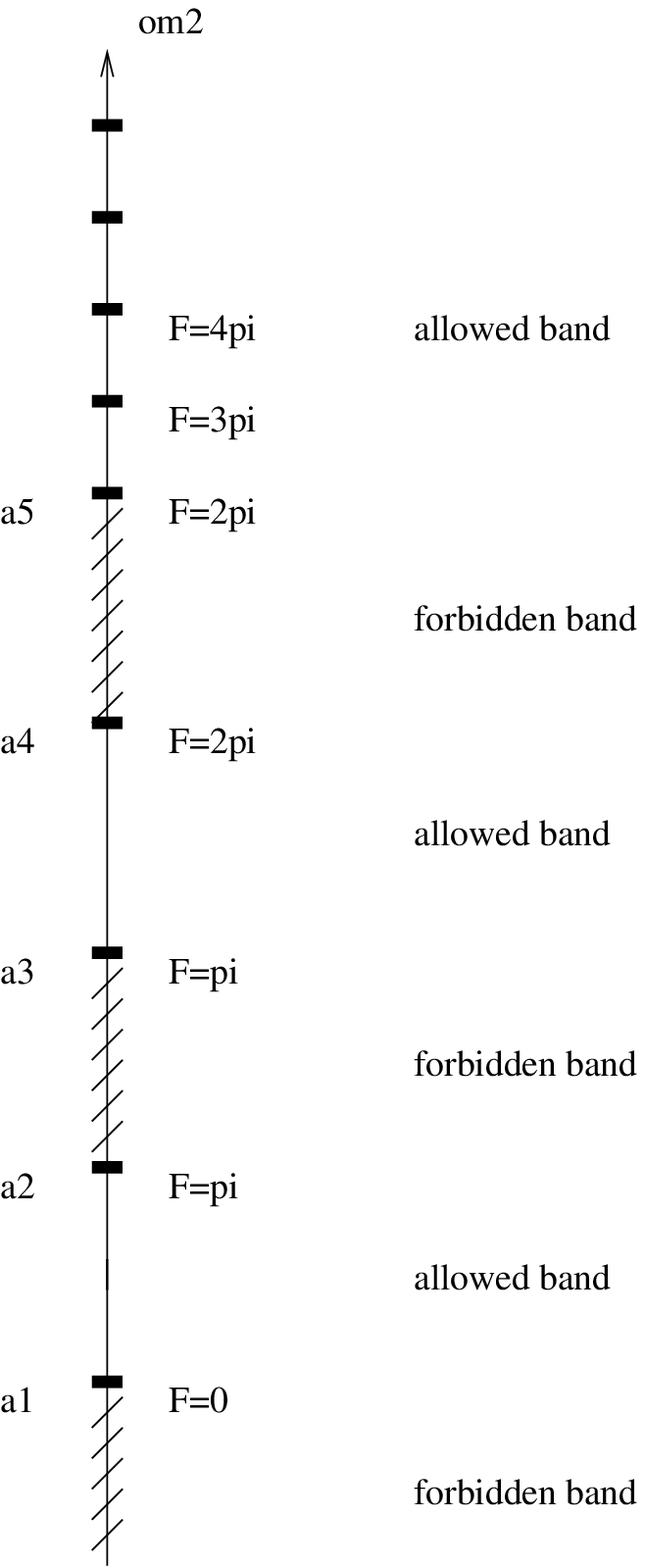,height=8cm,width=10cm}
\vspace{0.1cm} \caption{Spectrum of eq.\,(\ref{phi4semicl})}
\label{spectrumphi4}
\end{figure}

The stability equation (\ref{stability}) around the background
(\ref{phi4kink}) takes the form
\begin{equation}\label{phi4semicl}
\left\{\frac{d^{2}}{d \bar{x}^{2}}+\bar{\omega}^{2}+1-3\,
(2-\bar{\phi}_{0}^{2})\,\;\textrm{sn}^{2}\left(\frac{\bar{\phi}_{0}}{\sqrt{2}}\,\bar{x}\,,\,k^2\right)
\right\}\bar{\eta}(\bar{x})=0\;,
\end{equation}
where $\bar{\omega}=\omega/m$, and it can be reduced to the Lam\'e
equation with $N=2$ (see Appendix \ref{lame}). The allowed and
forbidden bands, with corresponding values of the Floquet
exponent, are shown in Fig. \ref{spectrumphi4}.

\psfrag{F=4pi}{$F=4\pi$}\psfrag{F=3pi}{$F=3\pi$}\psfrag{F=2pi}{$F=2\pi$}\psfrag{F=pi}{$F=\pi$}\psfrag{F=0}{$F=0$}
\psfrag{om2}{$\bar{\omega}^2$}\psfrag{allowed band}{allowed
band}\psfrag{forbidden band}{forbidden
band}\psfrag{a1}{\hspace{-1.5cm}\small$1-\frac{2\sqrt{k^4-k^2+1}}{1+k^2}$}
\psfrag{a2}{\hspace{0.5cm}\small$0$}\psfrag{a3}{\small$\frac{3
k^2}{1+k^2}$}\psfrag{a4}{\small$\frac{3}{1+k^2}$}
\psfrag{a5}{\hspace{-1.5cm}\small$1+\frac{2\sqrt{k^4-k^2+1}}{1+k^2}$}

The boundary conditions (\ref{antiperbc}) translate into the
requirement of antiperiodicity for the fluctuation $\eta$
$$
\eta(x+R)\,=\,-\,\eta(x)\,,
$$
which selects the values of $\bar{\omega}^{2}$ for which the
Floquet exponent is an odd multiple of $\pi$. These eigenvalues
are the zero mode
\begin{equation}\label{omega0}
\bar{\omega}_{0}^{2}=0\;,
\end{equation}
the discrete value
\begin{equation}\label{omegad}
\bar{\omega}_{1}^{2}=\frac{3 k^2}{1+k^2}\;,
\end{equation}
and the infinite series of points (with multiplicity $2$) inside
the highest band
\begin{equation}\label{omegan}
\bar{\omega}_{n}^{2}\equiv 1-\frac{3}{1+k^2}\left[{\cal P
}(a_{n})+{\cal P }(b_{n})\right]\;,
\end{equation}
with $a_{n},b_{n}$ constrained by
\begin{equation}\label{anbn}
\begin{cases}F=2
i\left\{\textbf{K}[\zeta(a_n)+\zeta(b_n)]-(a_n+b_n)\,\zeta(\textbf{K})\right\}=(2 n-1)\pi\\
{\cal P}'(a_n)+{\cal P}'(b_n)=0\end{cases},\qquad n=2,3,...
\end{equation}

In the IR limit ($k\rightarrow 1 $) this spectrum goes to the one
related to the standard background (\ref{phi4kinkinfvol}). In
fact, the allowed band
$\;1-\frac{2\sqrt{k^4-k^2+1}}{1+k^2}<\bar{\omega}^{2}<0\;$ shrinks
to the eigenvalue $\,\bar{\omega}_{0}^{2}=0\,$, the other band
$\;\frac{3 k^2}{1+k^2}<\bar{\omega}^{2}<\frac{3}{1+k^2}\;$ shrinks
to $\,\bar{\omega}_{1}^{2}=\frac{3}{2}\,$, and finally
$\;\bar{\omega}^{2}>1+\frac{2\sqrt{k^4-k^2+1}}{1+k^2}\;$ goes to
the continuous part of the spectrum
$\,\bar{\omega}_{q}^{2}=2+\frac{1}{2}q^{2}\,$.

In order to complete the spectrum also at values $mR<\pi$, we
have to put together the frequencies (\ref{omegad}) and (\ref{omegan})
with the ones obtained by quantizing the constant solution
(\ref{phi4kinkUV}). We therefore obtain\footnote{To be precise,
notice that the eigenvalue
$\bar{\omega}_{1}^{2}=-1+\frac{\pi^2}{m^2 R^2}$ is double, and at
$mR=\pi$ it splits into the two simple eigenvalues (\ref{omega0})
and (\ref{omegad}).}
\begin{equation}\label{omegadcomplete}
\bar{\omega}_{1}^{2}=\begin{cases}\frac{3
k^2}{1+k^2}&\qquad\text{for} \qquad mR>\pi\\
-1+\frac{\pi^2}{m^2 R^2}&\qquad\text{for} \qquad
mR<\pi\end{cases}\;,
\end{equation}
and
\begin{equation}\label{omegancomplete}
\bar{\omega}_{n}^{2}= \begin{cases}1-\frac{3}{1+k^2}\left[{\cal P
}(a_{n})+{\cal P }(b_{n})\right]
&\qquad\text{for} \qquad mR>\pi\\
-1+(2n-1)^2\frac{\pi^2}{m^2 R^2}&\qquad\text{for} \qquad
mR<\pi\end{cases}\;.
\end{equation}

\begin{figure}[ht]
\begin{center}
\psfig{figure=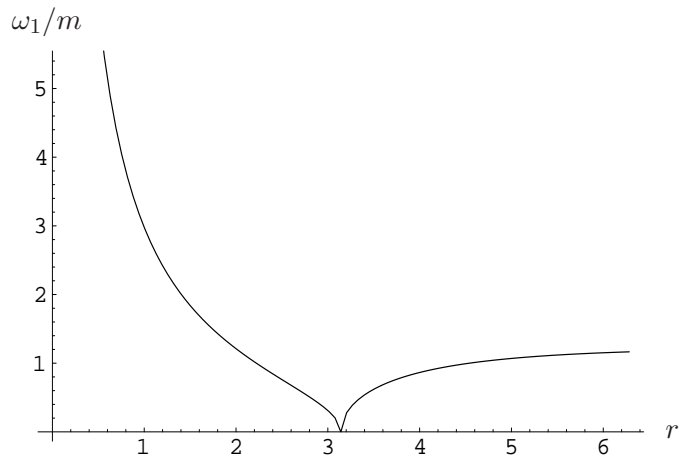,height=6cm,width=9cm} \caption{The first
level defined in (\ref{omegadcomplete})}\label{figomegad}
\end{center}
\end{figure}

With the explicit knowledge of the stability frequencies, and in
particular of the first one, plotted in Fig.\,\ref{figomegad}, we
can now understand the physical meaning of the point $mR=\pi$.
This corresponds, in fact, to the limit $k\to 0$ and this is the value
below which the analytic continuation of the classical background
(\ref{phi4kink}) becomes imaginary. Correspondingly, the first frequency square
$\omega_1^2$ tends to zero, and its continuation would become
negative, signaling an instability of the solution. At the same
time, the constant background (\ref{phi4kinkUV}) is stable just up
to the point $mR=\pi$, as it can be easily seen from
Fig.\,\ref{figomegad}.

\psfrag{omegad}{$\omega_{1}/m$}\psfrag{r}{$r$}

\begin{figure}[ht]
\begin{center}
\psfig{figure=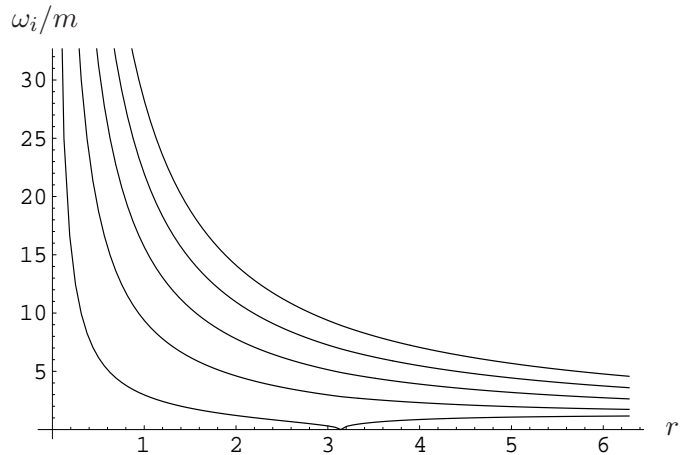,height=6cm,width=9cm} \caption{The first
few levels defined in (\ref{omegadcomplete}) and
(\ref{omegancomplete})}\label{figomegai}
\end{center}
\end{figure}

Figure \ref{figomegai} shows the plots, for generic values of $r$
in (\ref{scalvar}), of the first few frequencies given by
(\ref{omegadcomplete}) and (\ref{omegancomplete}), which represent
the energies of the excited states with respect to their ground
state $E_{0}(R)$.

\psfrag{omegad}{$\omega_{i}/m$}\psfrag{r}{$r$}

We have now all data to write the ground state energy
in the kink sector, which is defined in analogy with the infinite
volume case (\ref{grstateinfvol}) as
\begin{equation}\label{grstate}
E_0(R) \,=\, {\cal E}_{cl}(R) \,+\, \frac{1}{2}\,\sum_{i}
\omega_i(R) \,+\,C.T.\,-\, \frac{1}{2}\sum_{n=-\infty}^{\infty}
\omega_n^{\text{vac}}(R) \,\,\,,
\end{equation}
where the frequencies $\omega_i$ are defined in
(\ref{omegadcomplete}) and (\ref{omegancomplete}), and the mass
counterterm is defined as
$$
C.T.\,=\,-\, \frac{\delta m^{2}}{2}\int\limits_{-R/2}^{R/2}dx
\left[\left(\phi_{cl}^{\text{kink}}(x)\right)^2-\frac{m^2}{\lambda}\right]\;,
$$
with
$$
\delta
m^{2}\,=\,\left.\frac{3}{4\pi}\,\lambda\,\frac{2\pi}{R}\,\sum_{n=-\infty}^{\infty}
\frac{1}{\omega_n^{\text{vac}}}\right.\;,\qquad\qquad
\omega_n^{\text{vac}}(R)=\sqrt{2 m^2+\left(\frac{2 n
\pi}{R}\right)^2}\;.
$$
A more transparent expression for the ground state energy
(\ref{grstate}), which explicitly shows the cancellation of the
divergencies present in each term separately, can be obtained by
expanding all quantities around some specific value of $r$. In
particular, in the limits of large or small $r$ one can extract
the asymptotic IR and UV data of the theory. We have already seen
in Sect.\,\ref{sectclasssol} how the large-$r$ expansion of the
classical energy correctly encodes the scattering data of the
infinite volume theory, and we will now study the UV limit $r\to 0
$, in which we can extract some conformal data related to the
theory in exam. Furthermore, in Appendix\,\ref{appfinite} we
perform the expansion around the point $r=\pi$, where it is
possible to see how the divergencies cancel in a more subtle way.

The small--$r$ expansion of (\ref{grstate}) is easily obtained to
be
\begin{equation}\label{grstexp}
\frac{E_0(R)}{m} \,=\,
\frac{2\pi}{r}\left[\sum\limits_{n=1}^{\infty}\left(n-\frac{1}{2}\right)-\sum\limits_{n=1}^{\infty}n\right]
\,-\,\frac{1}{4\sqrt{2}}\,+\,\frac{r}{2\pi}\left[\frac{\pi}{2}\,\frac{m^2}{\lambda}-
\sum\limits_{n=1}^{\infty}\frac{1}{2n-1}+
\sum\limits_{n=1}^{\infty}\frac{1}{2n}\right] \,+\,...\;.
\end{equation}
The individually divergent series present in (\ref{grstexp})
combine to give a finite result, in virtue of the relations
\begin{eqnarray*}
&&\sum\limits_{n=1}^{\infty}
(2n-1)\,-\,\sum\limits_{n=1}^{\infty}(2n)\,=\,2
\left[\zeta(-1,1/2)-\zeta(-1)\right]\,=\,2\left[\frac{1}{24}
+\frac{1}{12}\right]\;,\\
&&
\sum\limits_{n=1}^{\infty}\frac{1}{(2n-1)}\,-\,\sum\limits_{n=1}^{\infty}\frac{1}{(2n)}\,=\,
\sum\limits_{k=1}^{\infty}\frac{(-1)^{k+1}}{k}\,=\,\log 2\;.
\end{eqnarray*}

The UV behaviour for $r \rightarrow 0$ of the ground state energy
$E_0(R)$ of a given off--critical theory is related to the
Conformal Field Theory (CFT) data $(h,\bar{h},c)$ of the
corresponding critical theory and to the bulk energy term as
\begin{equation}
E_0(R) \,\simeq\,\frac{2\pi}{R} \left(h + \bar{h} -
\frac{c}{12}\right) + {\cal B} \, R + \cdots\; \label{UVCFT}
\end{equation}
where $c$ is the central charge, $h + \bar{h}$ is the lowest
anomalous dimension in a given sector of the theory and ${\cal B}$
the bulk coefficient. Therefore, we estimate the semiclassical
bulk term to be given by
$$
{\cal
B}\,=\,m^2\left(\frac{1}{4}\,\frac{m^2}{\lambda}\,-\,\frac{\log
2}{2\pi}\right)\;.
$$
Furthermore, our result for the leading semiclassical term in the
anomalous dimension is \footnote{Notice that the central charge
contribution $-c/12$ is absent in (\ref{grstexp}), because we are
subtracting the ground state energies of kink and vacuum sector,
which both have the same central charge $c=1$.}
\begin{equation}
h+\bar{h}\,=\,\frac{1}{8}\;.
\end{equation}
As it is fully discussed in Sect.\,\ref{sectkinkop}, this result
agrees with the CFT prediction.

Finally, again in accordance with the CFT expectation, the excited
levels are given by
\begin{equation}
E_{k_n}(R) \,\simeq\,\frac{2\pi}{R}
\left[\frac{1}{8}+\sum\limits_{n}\,k_n\left(n-\frac{1}{2}\right)\right]
+ \left[{\cal
B}-\frac{m^2}{2\pi}\sum\limits_{n}\frac{k_n}{2n-1}\right] \, R +
\cdots\; \label{UVCFT}
\end{equation}

\subsection{Other interpretations of the classical solution}


In concluding this Section, it is worth to comment how the
classical solution (\ref{phi4kink}) has been studied in the
literature either in different contexts, or in the same as ours
but along different lines of interpretation.

In fact, this kind of background, regarded however as a
time--dependent solution in zero space dimensions, has been
proposed in \cite{harrington} to describe dominating contributions
to the partition function at finite temperature $T$, i.e. when the
Euclidean time variable is compactified on a circumference
$\beta=\frac{1}{k_B T}$ with \textit{periodic} b.c.. In this case,
the finite value of $T$ which corresponds to $k=0$ is naturally
interpreted as a limiting temperature, above which no periodic
solutions exist.

Moreover, the background (\ref{phi4kink}) has also been studied in
\cite{manton,liang} as a static classical solution on a
cylindrical geometry. In these works, however, \textit{periodic}
b.c. are considered, and the size of the system is related to the
elliptic modulus as
$$
mR=\sqrt{1+k^2}\;\,4 \,N\,\textbf{K}(k^{2})\;,
\qquad\text{with}\quad N\in\mathbb{N}\;.
$$
This choice, which corresponds to considering the solution as a
train of $N$ kinks and $N$ antikinks, implies the selection of $N$
distinct eigenvalues with $\omega_n^2<0$ in the spectrum of
eq.\,(\ref{phi4semicl}). Their imaginary contributions to the
energy levels indicate the instability of the considered
background, which is explained in \cite{liang} by noting that in
the $k\to 1$ ($R\to\infty$) limit the solution tends to a single
kink, instead of keeping its periodic nature of a train of kinks
and antikinks. All this reflects the ambiguity present in the
definition of the size of the system $R$ in terms of the elliptic
modulus $k$, simply due to the periodicity of the Jacobi function
$\text{sn}(u,k)$, and correspondingly in the interpretation of the
solution for a chosen definition of $R$.
However, choosing (\ref{sizephi4}), i.e. \textit{antiperiodic}
b.c., the infinite volume limit is smoothly recovered as $k\to 1$,
and the corresponding single kink solution is stable. It is then
natural to expect that for any value of $R$ of the finite system,
also time--dependent solution exist, which describe multikink or
kink--antikink configurations. Such solutions can be quantized in
finite volume as well, although this is a subject that is out
of the scope of the present paper.

Finally, in the recent paper \cite{cho} the orbifold geometry
$S^{1}/\mathbb{Z}_2$ is considered, instead of the circle, for the
worldsheet space coordinate $x$, and a classical background very
similar to (\ref{phi4kinkcomplete}) is introduced. The analogy
with our case, however, is only apparent. In fact, due to the
absence of translational invariance, on the orbifold the kink and
the antikink have to be considered as two distinct degenerate
solutions, suggesting therefore a phase transition at $mR=\pi$. In
our case, on the contrary, the lowest energy level is never
degenerate, consistently with the fact that the behavior of the
scaling functions at $mR=\pi$ does not hint at any underlying
conformal field theory. The discontinuity of the derivative of
$\omega_1$ at $mR=\pi$ should be then interpreted as just an
effect of the semiclassical approximation.

\section{Kink--creating operators in Landau--Ginsburg models}
\label{sectkinkop}\setcounter{equation}{0}

As it is well known, starting from $c=1$  CFT in two dimensions
and adding to its Lagrangian different relevant operators with an
appropriate choice of the coupling constants, one can construct
many integrable and non-integrable 2D massive QFT's having
degenerate vacua \cite{LG}. They can be classified according to
the symmetries preserved by the perturbation. For instance, SG and
Double SG models are examples of $Z \otimes Z_2$--invariant
theories (i.e. $\phi \to \pm \phi+2\pi n$ ), while LG models of
$Z_2$--invariant (i.e. $\phi \to -\phi$) ones. The common feature
of all these models are the non--perturbative topologically stable
classical solutions (solitons or kinks) interpolating between two
vacua. In the quantum theory they give rise to specific \lq\lq
strong coupling" particles, carrying topological ($Z$ or/and
$Z_2$) charges and representing an important part of their IR
spectrum. The description of the finite volume spectrum (on the
cylinder) of these models therefore requires both the construction
of the finite volume counterparts of such topological solutions
and the identification of the quantum states related to them. An
important consistency check for the finite volume spectrum is
provided by its UV and IR limits (in the scaling variable $mR$)
that should reproduce the CFT and the massive model spectra
correspondingly. In order to understand the flow between the UV
theory to the IR one, i.e. the relation between the CFT space of
states (and the corresponding field operators) and the infinite
volume (massive) particle space of states, it is also necessary to
recognize the states (and operators) that describe  the UV limits
of  such solitons and kinks in the $c=1$ CFT. The soliton (and
kink) creating operators are non-local functionals of the  field
$\phi$ that satisfy the following requirements:

(a) to carry ($Z$ or $Z_2$) topological charges $\pm 1$ or
equivalently to produce specific b.c.'s \footnote{the relation
between the $z$ and $\bar z$ coordinates used in this section and
the $x$ and $t$ used in  all  the others is the standard plane to
cylinder one, i.e. $z=e^{\frac {i} {R}(x+t)}$ and $\bar
z=e^{-\frac {i} {R}(x - t)}$.} for $\phi$,
\begin{equation}\label{bcsol}
\phi(ze^{2i\pi},{\bar z}e^{-2i\pi})=\phi(z,\bar z) +2\pi n {\cal
R} \qquad n=\pm 1
\end{equation}
for solitons (where ${\cal R}$ is the compactification radius of
$\phi$, say  ${\cal R}=\beta ^{-1}$ for SG), and
\begin{equation}\label{bcskink}
\phi(ze^{2i\pi},{\bar z}e^{-2i\pi})=-\phi(z,\bar z)
\end{equation}
for the ($Z_2$) kinks.

(b) to be local with respect to the perturbation (i.e.,
$V_l(\phi)\,=\,\sum\limits_{k=1}^{l}\,\lambda_k\phi^{2k-2}$ for
the LG models) or/and to the corresponding energy density operator
in order to have well defined off-critical properties.

Before discussing  the construction of the kink-creating operators
for the LG models (\ref{LGpot}), it is worthwhile to remind how the
soliton operators are derived in the case of SG model \cite{mand,
BerLec}. As it well known \cite{zzreview}, the primary fields in
the untwisted (\lq\lq winding") sector\footnote{defined by the
condition that the chiral U(1) currents $I(z)= {\partial
\varphi}(z) $ and $\bar I(\bar z)={\bar
\partial}\varphi(\bar z)$ are single valued.} of the (compact)
$c=1$ Gaussian CFT are represented by the following {\em discrete} set
of vertex operators
$$V_{n,s}(z,\bar z)=:\exp( ip\phi + i\bar
p\tilde {\phi}):$$
with
$$ p={{s} \over {\cal R}}
\,\,\,\,\,\,
,
\,\,\,\,\,\,
\bar p = 2\pi
gn {\cal R}
\,\,\,\,\,\,
,
\,\,\,\,\,\,
n,s=0,\pm 1,\pm 2,...
$$
Their \lq\lq chiral"
dimensions\footnote{Note that we have introduced arbitrary
normalization constant $g$ in the action $A_{gauss}
\,=\,\frac{g}{2}\int d^2 x \left(\partial_\mu
\phi\right)\left(\partial^\mu \phi\right)$ and as a consequence
the the chiral component of the stress-tensor $T(z,\bar z)$ is
given by $T = -2\pi g:(\partial \phi)^2:$. The standard CFT
normalization is $g=\frac{1}{2\pi}$, but we shall often use
$g=1$.} are given by $ h = {{(p+\bar p)^2} \over {8\pi g}}$ and $h
= {{(p-\bar p)^2} \over {8\pi g}}$ and therefore they have spin
$s=h - \bar h$ and dimension $ \Delta = h + \bar h $. We have
introduced the free fields $\varphi(z)$ and $\bar {\varphi}(\bar
{z})$ such that $\phi= \varphi(z) + \bar {\varphi}(\bar {z})$ and
its dual is $\tilde \phi= \varphi(z) - \bar {\varphi}(\bar {z})$.
They  take values on the circle $S_1$ with radius ${\cal R}=\frac
{1}{\beta}$ and their correlation functions have the form:
\begin{equation}\label{corrfts}
<\varphi(z)\varphi(w)>\,=\, -{{1} \over {4\pi g}} \ln(z-w)
,\qquad\qquad <\bar \varphi(\bar {z})\bar\varphi(\bar {w})>=- {{1}
\over {4\pi g}} \ln(\bar {z}-\bar {w})
\end{equation}
As one can easily verify from the OPE
\begin{equation}\label{ope}
\phi(z,\bar z) V_{n,s}(0,0) = -{{i} \over {4\pi g}} \left({\bar
p}\, \ln\left({z \over \bar z}\right) + p \,\ln\left(z \bar
z\right)\right) V_{n,s}(0,0) + ...
\end{equation}
the vertex operators $V_{n,s}$ for $n=\pm 1$ and  for arbitrary
spin s, create the $Z$--type b.c.'s (\ref{bcsol}) (in fact one can
take, say $s=0$ or $s=\pm 1$, since the only $\tilde \phi$
contribution is relevant). They are also local with respect to the
SG potential $V_{SG}={{m^2}\over {\beta^2}}\cos(\beta \phi)$ as it
follows from their OPE's (with ${\cal R}=\frac {1}{\beta}$)
\begin{equation}\label{opecos}
\cos(\beta \phi(z,\bar z)) V_{n,s}(0,0) ={1 \over 2} \left({z
\over \bar z}\right)^{{\bar p\beta} \over {4\pi g}}(z \bar z)^{{ p
\beta} \over {4\pi g}} V_{n,s+1}(0,0) + {1 \over 2}\left({z \over
\bar z}\right)^{-{{\bar p\beta} \over {4\pi g}}}(z \bar z)^{-{{ p
\beta} \over {4\pi g}}} V_{n,s-1}(0,0) + ...
\end{equation}
i.e. we have no changes under the transformation
$(ze^{2i\pi},{\bar z}e^{-2i\pi})$ to $(z,\bar z)$, since $ \bar p
= \frac{2\pi gn}{\beta}$ and ${{\bar p\beta} \over {4\pi g}} =
\frac {n}{2}$. Therefore  for $n = \pm1$ they represent the one
soliton--creating operators. The operators with $n\geq 2$ create
multi--soliton states. It should be noted that in the perturbed
CFT (i.e. in  SG theory) the dual field $\tilde \phi$ is nonlocal
in terms of the SG field $\phi$, i.e. we have $\tilde \phi(x,t) =
\int\limits_{-\infty}^{x}dy \partial_y \phi(x,y)$. The $Z$
topological (i.e. soliton) charge $Q$ is defined by the
eigenvalues of the well known SG charge operator
\begin{equation}\label{topcharge}
Q \,=\, {{\beta}\over {2\pi}} \int\limits_{-\infty}^{\infty}dx
\partial_x \phi(x,t)\;.
\end{equation}

In order to describe the operators that create  $Z_2$--type
(antiperiodic) b.c.'s (\ref{bcskink}) for the SG field $\phi$ we
have to consider the twisted sector of the $c=1$ CFT. It is
defined (see ref. \cite {zamtwist}) by the condition that the
chiral $U(1)$ currents $I(z) = {\partial \varphi}(z) $ and $\bar
I(\bar z)={\bar \partial}\varphi(\bar z)$ are double valued, i.e.
their mode expansions contain only half--integer modes
\begin{equation}\label{modex}
I(z) \,=\,\sum\limits_{m=-\infty}^{\infty}\, I_{m-{1\over
2}}\,z^{-m-{1\over 2}},\qquad\qquad \bar I(\bar z)
\,=\,\sum\limits_{m=-\infty}^{\infty}\,{\bar I}_{m-{1\over
2}}\,{\bar z}^{-m-{1\over 2}}
\end{equation}
where the modes $I_{m-{1 \over 2}}$ (and $ {\bar I}_{m-{1 \over
2}}$ ) satisfy the following Heisenberg type algebra:
\begin{equation}\label{algI}
[ I_{m-{1\over 2}},I_{l-{1\over 2}}] = {{m-{1\over 2}} \over 2}
\delta_{m+l},\qquad\qquad [ I_{m-{1\over 2}}, {\bar I}_{l-{1\over
2}}] = 0
\end{equation}
The primary fields in this sector  ${\mu}_{k,\bar k}^{\pm}$, i.e.
\begin{equation}\label{primary}
 I_{m + {1\over 2}} {\mu_{k,\bar k}^{\pm} }= 0 \qquad\qquad {\bar I}_{m + {1\over 2}} {\mu_{k,\bar k}^{\pm}} = 0,
\qquad\qquad m,\bar k, k = 0,1,2,...
\end{equation}
have \lq\lq chiral" dimensions $ h_k = {{(2k+1)^2}\over {16}}$
and $ \bar h_{\bar k} =  {{(2\bar k+1)^2}\over {16}}$ and the
allowed spins
 are given by $s=0,\pm{1 \over 2}$. As one can see from
the OPE
\begin{equation}\label{opemu}
\phi(z,\bar z) \mu_0^{\pm}(0,0) = \sqrt {z} \mu_1^{\pm}(0,0) +
\sqrt {\bar z} \bar{\mu}_1^{\pm}(0,0) +...
\end{equation}
the fields $\mu_{0,0}^{\pm}(0,0)=\mu_0^{\pm}$ (of  lowest
dimension $ h + \bar h = {1 \over 8}$ and spin $s = 0$), called
disorder (or spin) fields, create branch cut singularity for
$\phi$ and thus reproduces the $Z_2$--type b.c.'s (\ref{bcskink}).
Their locality with respect to $\cos(\beta \phi)$ is a consequence
of the OPE (\ref{opemu}) and of the following correlation function
\begin{eqnarray}\nonumber
& <\mu_0^{-}(\infty,\infty)e^{i\alpha \phi(w,\bar w)} \cos(\beta \phi(z,\bar z)) \mu_0^{+}(0,0)> \,=\\
& =\,{{C_{+-}} \over {2}} \left[\left({{(\sqrt {w} - \sqrt
{z})(\sqrt {\bar w} - \sqrt {\bar z})} \over {(\sqrt {w} + \sqrt
{z})(\sqrt {\bar w} + \sqrt {\bar z})}} \right) ^{{\alpha \beta}
\over {4\pi g}}
 +\left({{(\sqrt {w} - \sqrt  {z})(\sqrt {\bar w} - \sqrt {\bar z})} \over {(\sqrt {w} + \sqrt {z})(\sqrt {\bar w} +
 \sqrt {\bar z})}} \right)
 ^{-{{\alpha \beta} \over {4\pi g}}}\right]\;.
\end{eqnarray}
Note that the current $I(z)$ does not have zero mode in the
twisted sector and therefore the fields ${\mu}_{k,\bar k}^{\pm}$
do not carry $U(1)$ (and $Z$), but only $Z_2$ charge. All these
properties of  the disorder field $\mu_0^{\pm}(0,0)$ lead to the
conclusion  that it represents the kink--creating operator. It
should be mentioned that the field $\phi$ in this case takes its
values on the orbifold ${{S_1} \over {Z_2}}$ and, as usually, the
two disorder fields $\mu_0^{\pm}(0,0)$ are related to the two
fixed points  $\phi = 0$ and $\phi = \pi {\cal R}$ (\cite{dixon}).
As we shall show in Sect.\,\ref{sectSG}, in finite volume one can
have both the quasiperiodic (soliton type) and antiperiodic (kink
type) solutions and states, which however in the IR (infinite
volume) SG theory are related to the  same soliton (and
anti-soliton) states.

The description of the kink--creating operators in the LG models
is quite similar to the one of the SG model. The main difference
is that the field $\phi$ is no longer compactified, i.e. it lives
now on the orbifolded line ${R^{(1)} \over {Z_2}}$. The untwisted
(i.e. $Z_2$--even) sector of the corresponding (noncompact) $c=1$
CFT contains  two {\em continuous} parameters $(q,\bar q)$ family
of vertex operators $V_{q,\bar q} = :\exp(i q \phi + i \bar q
\tilde {\phi}):$ of \lq\lq chiral" dimensions $ h = {{(q+\bar
q)^2} \over {8\pi g}}$ and $h = {{(q-\bar q)^2} \over {8\pi g}}$.
As in SG case the operators with $\bar q \neq 0$ produce certain
nontrivial b.c.'s for $\phi$, but with continuous $U(1)$ charge.
As expected, there is not a properly defined $Z$ topological
charge in this case. Such operators are also non--local with
respect to the LG potential (\ref{LGpot}) as it can be seen from
the OPE's, say
\begin{equation}\label{opephi}
:{\phi(z,\bar z)}^k: :e^{i\bar q \tilde \phi(0,0)}: \,=\, :\left(-
{{i\bar q} \over {4\pi g}} ln{{z} \over {\bar z}} +
\phi(0,0)\right)^k e^{i\bar q \tilde \phi(0,0)}: +...
\end{equation}
Therefore they cannot represent kink--creating operators. The
structure of the twisted sector of this noncompact $c=1$ CFT is
quite similar to the one considered in the context of the SG (i.e.
$\cos(\beta \phi)$ ) perturbation above. Since in the orbifold
line (as well as in orbifold finite interval) we have only one
fixed point $\phi =0$, we have correspondingly only one disorder
field $\mu_0 $ of dimension $1/8$ and spin zero. As in the SG
case, the field $\mu_0$  produces branch cut in the OPE with $\phi$
and so, it implements the $Z_2$--type (antiperiodic) b.c.'s (\ref{bcskink}).
In order to check whether it is local with respect to the LG
potential let's consider its correlation functions
\begin{eqnarray}\label{phi}
&<\mu_0(\infty,\infty)e^{i\alpha \phi(w,\bar w)} e^{i\gamma
\phi(z,\bar z)}  \mu_0(0,0)> \, = \, C_0 \left({{\sqrt {w} - \sqrt
{z}} \over {\sqrt {w} + \sqrt {z}}} \right)^{{\alpha \gamma}\over
{4\pi g}}
 \left({{\sqrt {\bar w} - \sqrt {\bar z}} \over {\sqrt {\bar w} + \sqrt {\bar z}}} \right)
 ^{{\alpha \gamma}\over {4\pi g}},\\
&<\mu_0(\infty,\infty)e^{i\alpha \phi(w,\bar w)}  :{\phi(z,\bar
z)}^k : \mu_0(0,0)> \,=\,
 C_0{(-i)}^k \left({{ \alpha} \over {4\pi g}} \ln ({{\sqrt {w} - \sqrt  {z}} \over {\sqrt {w} + \sqrt {z}}} )
 ({{\sqrt {\bar w} - \sqrt {\bar z}} \over {\sqrt {\bar w} + \sqrt {\bar z}}} )\right)^k
\end{eqnarray}
These can be derived from the $\varphi$ mode expansion $\varphi(z)
\,=\,\sum\limits_{m=-\infty}^{\infty}\, {{I_{m-{1\over 2}}} \over
{{1 \over 2}-m}}\, z^{-m+{1\over 2}}$, the algebra (\ref{algI}) of
its modes and the properties (\ref{primary}) of the disorder field
$\mu_0$. It is now easy to see that each (linear) combination of
even powers of the field $\phi$ is local with respect to $\mu_0$,
i.e. it does not change under the transformation $(ze^{2i\pi},{\bar z}e^{-2i\pi})$
to $(z,\bar z)$. It becomes clear from this discussion that the
only field that can create $Z_2$--kinks in the LG models is then the
disorder field $\mu_0$. In the \lq\lq broken phase" $\phi^4$ model
(\ref{phi4pot}) we have only one kink interpolating between the
two minima of the potential. In the symmetric type LG
potentials, as for example
\begin{eqnarray}\nonumber
V_l^{\text{odd}}&=&\frac{1}{2}\,\prod\limits_{k=1}^{\frac{l-1}{2}}\,\left(\phi^2-a_k^2\right)^2
\qquad \quad \text{for}\quad l=3,5,...\\
V_l^{\text{even}}&=&\frac{1}{2}\,\phi^2\,\prod\limits_{k=1}^{\frac{l-2}{2}}\,\left(\phi^2-a_k^2\right)^2
\qquad \text{for}\quad l=4,6,... \label{LGpot2}
\end{eqnarray}
we have instead a finite number of $l$ degenerate vacua and therefore
different kinks relating  each two consecutive vacua. An important
question is: how to distinguish them in a finite volume? Moreover,
in the CFT language, what are the operators which create such kinks?

To answer such questions, observe that the minima of these potentials
are at the points $\phi_k =\pm a_k$
($k=1,2,...{{l-1} \over {2}}$ for $l$ odd) and since we consider
$a_1 > a_2>...$ the kinks are interpolating between $\phi_1$ and
$\phi_2$ ,etc. and not, as in the $\phi^4$ case, between $\pm
\phi_0$. Therefore the analog of the antiperiodic b.c.'s
(\ref{bcskink}) for the case of many degenerate vacua is given by
\begin{equation}\label{bcsLGkink}
\phi(ze^{2i\pi},{\bar z}e^{-2i\pi})= a_k +a_{k+1} - \phi(z,\bar
z),
\end{equation}
i.e. we have different b.c.'s for each kink. Indeed one can reduce
such b.c.'s to the standard ones (\ref{bcskink}) by introducing
the \lq\lq shifted" fields and the analog of the antiperiodic b.c.'s
(\ref{bcskink}) in the case of many degenerate vacua is given by
\begin{equation}\label{bcPhi}
\Phi_k(z,\bar z) = \phi(z, \bar z) - {(a_k +a_{k+1}) \over 2}
\end{equation}
In this scheme, however, the new fields have different vacua
expectation values. Since (different) orbifolds based on (\ref{bcsLGkink})
have different fixed points, one can formally prescribe to each such
point one disorder field $\mu^{(k)}(z,\bar z)$. As we shall see on
the example of the $\phi^6$ model in Sect.\,\ref{sectphiN} below,
although all these kinks have coinciding UV data, their finite
volume scaling functions are however different, with different
bulk coefficients etc.

\section{Sine-Gordon model with antiperiodic b.c.}
\setcounter{equation}{0}\label{sectSG}

In the light of the discussion of kink--creating operators
presented in Sect.\,\ref{sectkinkop}, it is worth to illustrate in
more detail the interesting case of the Sine--Gordon model, where
both kinds of kink exist. This fact can be easily understood in
the framework of the correspondence between Sine--Gordon and
Thirring models. In fact, the Sine--Gordon solitons are identified
with the Thirring fermions, for which two types of boundary
conditions (periodic and antiperiodic) can be naturally imposed in
a finite volume.

\begin{figure}[h]
\begin{tabular}{p{8cm}p{8cm}}

\footnotesize

\psfrag{phicl(x)}{$\beta\phi^{+}_{cl}$} \psfrag{2 pi}{$2\pi$}
\psfrag{K(k^2)}{$\textbf{K}(k^{2})$}
\psfrag{-K(k^2)}{$-\textbf{K}(k^{2})$}
\psfrag{x}{$\hspace{0.2cm}u^{+}$}

\psfig{figure=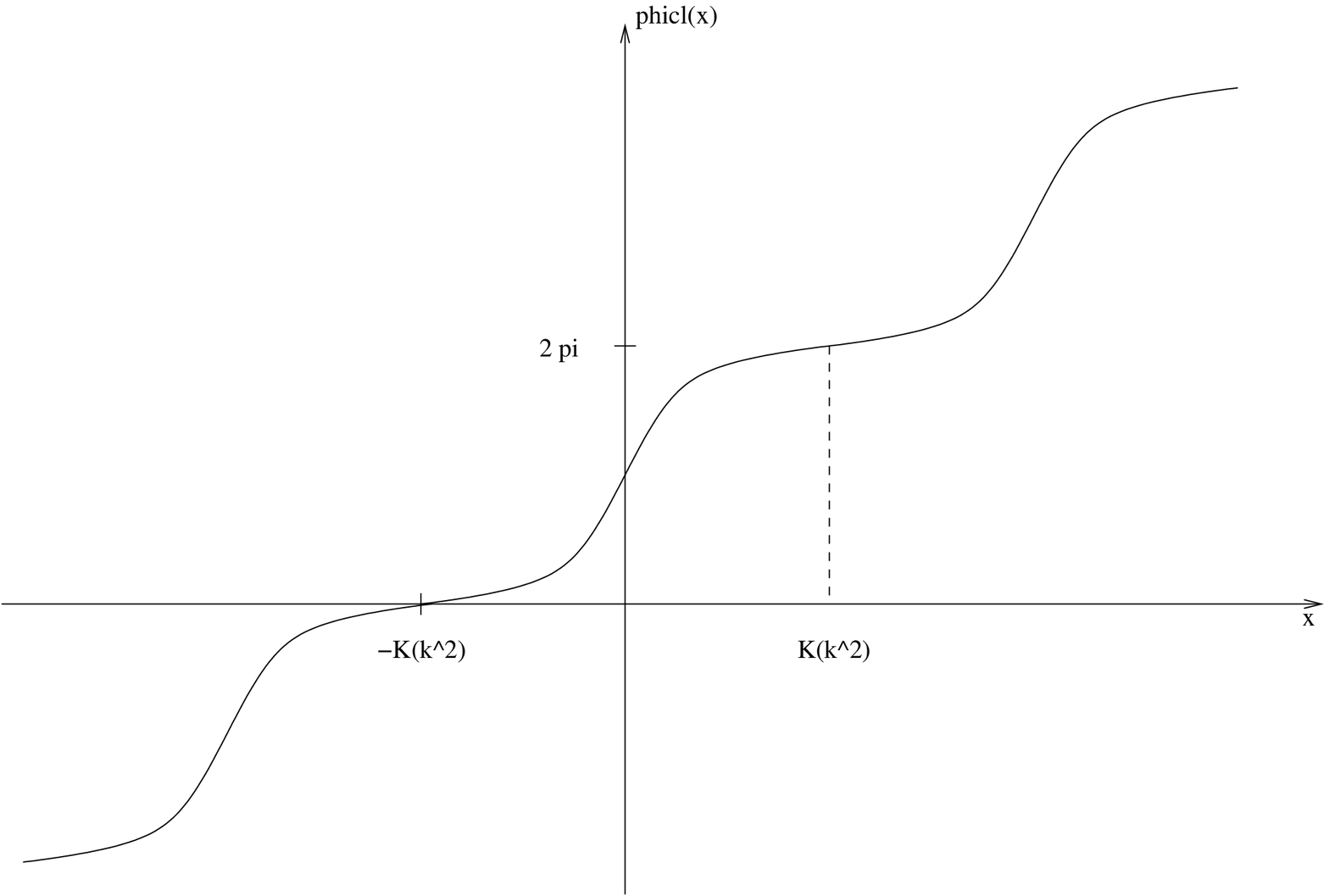,height=5cm,width=7cm}&

\footnotesize

\psfrag{phicl(x)}{$\beta\phi^{-}_{cl}$}
\psfrag{phi0}{$\beta\tilde{\phi}$} \psfrag{2
pi-phi0}{$2\pi-\beta\tilde{\phi}$}
\psfrag{K(k^2)}{$\textbf{K}(k^{2})$}
\psfrag{-K(k^2)}{$-\textbf{K}(k^{2})$}
\psfrag{x}{$\hspace{0.2cm}u^{-}$}

\psfig{figure=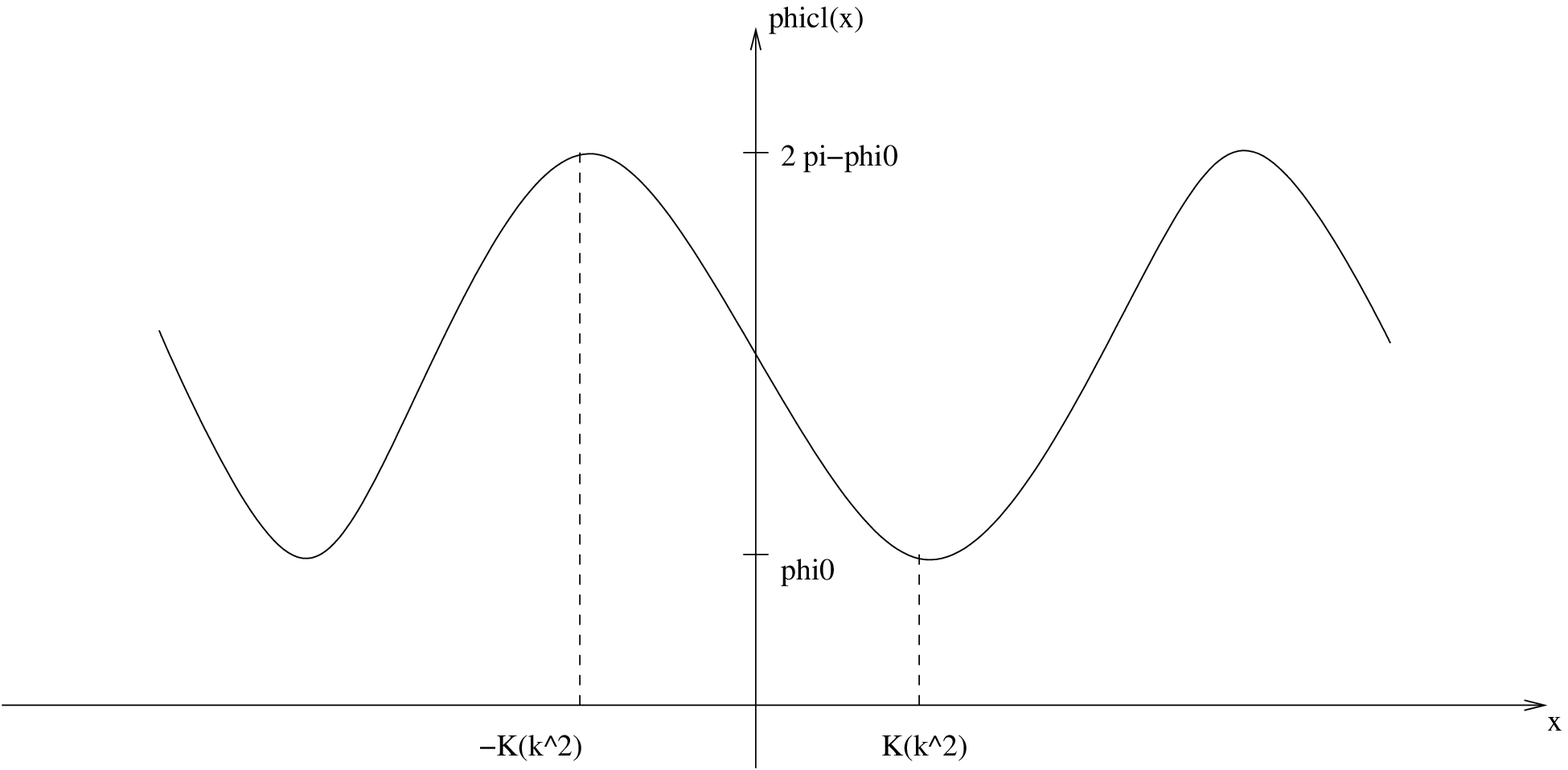,height=5cm,width=7cm}
\end{tabular}
\caption{Solutions of eq.\,(\ref{firstSG}), $A > 0$ (left hand
side), $-2 < A < 0$ (right hand side).}
 \label{figSGsol}
\end{figure}

The Euler--Lagrange equation for static backgrounds in the
Sine--Gordon model take the form
\begin{equation}
\label{firstSG} \frac{1}{2}\left(\frac{\partial
\phi_{cl}}{\partial x}\right)^{2} =
\frac{m^{2}}{\beta^{2}}\left(1-\cos\beta\phi_{cl} + A\right)
\,\,\,,
\end{equation}
and it admits three kinds of solution, depending on the sign of
the constant $A$. The simplest corresponds to $A=0$ and it
describes the standard kink in infinite volume:
\begin{equation}\label{SGkinkinfvol}
\phi^{0}_{cl}(x)\,=\,\frac{4}{\beta}\,\arctan\,e^{ m(x-x_{0})}\;.
\end{equation}
The other two solutions, relative to the case $A\neq 0$, can be
expressed in terms of Jacobi elliptic functions \cite{takoka},
defined in Appendix\,\ref{appell}. In particular, for $ A > 0$ we
have
\begin{equation}
\phi^{+}_{cl}(x)\,=\,\frac{\pi}{\beta} + \frac{2}{\beta}\,
\textrm{am}\left(\frac{ m (x - x_0)}{k},k\right)\;, \qquad k^{2}
\,=\,\frac{2}{2+A}\,\,\,, \label{SGam}
\end{equation}
which has the monotonic and unbounded behaviour in terms of the
real variable $u^{+}=\frac{ m (x - x_0)}{k}$ shown in
Fig.\,\ref{figSGsol}. For $ -2<A < 0$, the solution is given
instead by
\begin{equation}
\phi^{-}_{cl}(x)\,=\, \frac{2}{\beta}\,
\arccos\left[k\;\textrm{sn}\left( m (x - x_0),k\right)\right]\;,
\qquad k^{2} \,=\,1+\frac{A}{2}\,\,\,, \label{SGsn}
\end{equation}
and it oscillates in the real variable $u^{-}= m (x - x_0)$
between the $k$-dependent values $\tilde{\phi}$ and
$\frac{2\pi}{\beta}-\tilde{\phi}$ (see Fig.\,\ref{figSGsol}).

\vspace{0.5cm}

\vspace{0.5cm}

The solution (\ref{SGam}) satisfies quasiperiodic b.c.
\begin{equation}\label{quasiperbcSG}
\phi(x+R)\,=\,\phi(x)+\frac{2\pi}{\beta}\;,
\end{equation}
provided the circumference $R$ of the cylinder is identified with
$R=\frac{1}{m}\,2 \,k\,\textbf{K}\left(k^{2}\right)\;. $ The
complete semiclassical quantization of this background has been
performed in \cite{SGscaling}. It is worth to recall here the UV
limit of the corresponding energy levels, given by
\begin{eqnarray}\label{eiexpSG}
\frac{{\cal E}_{\{k_{n}\}}(R)}{m} & = &
\frac{2\pi}{r}\,\left(\frac{\pi}{\beta^{2}} +
\sum\limits_{n}k_{n}\,n\right)
\,-\,\frac{1}{4}\,+\,\frac{1}{\beta^{2}}\,r
\,-\,\frac{1}{8}\,\left(\frac{r}{2\pi}\right)^{2}\,+\, \\
&& \hspace{-3mm} -\, \left(\frac{r}{2\pi}\right)^{3}
\left[\frac{1}{8} \zeta(3) -\frac{1}{4}(2 \,\log 2 -1)
-\frac{\pi}{2\beta^{2}} +
\sum\limits_{n}k_{n}\,\frac{n}{4n^{2}-1}\right] + \ldots \nonumber
\end{eqnarray}
where $\{k_{n}\}$ is a set of integers defining a particular
excited state of the kink.

\begin{figure}[ht]
\begin{center}
\psfig{figure=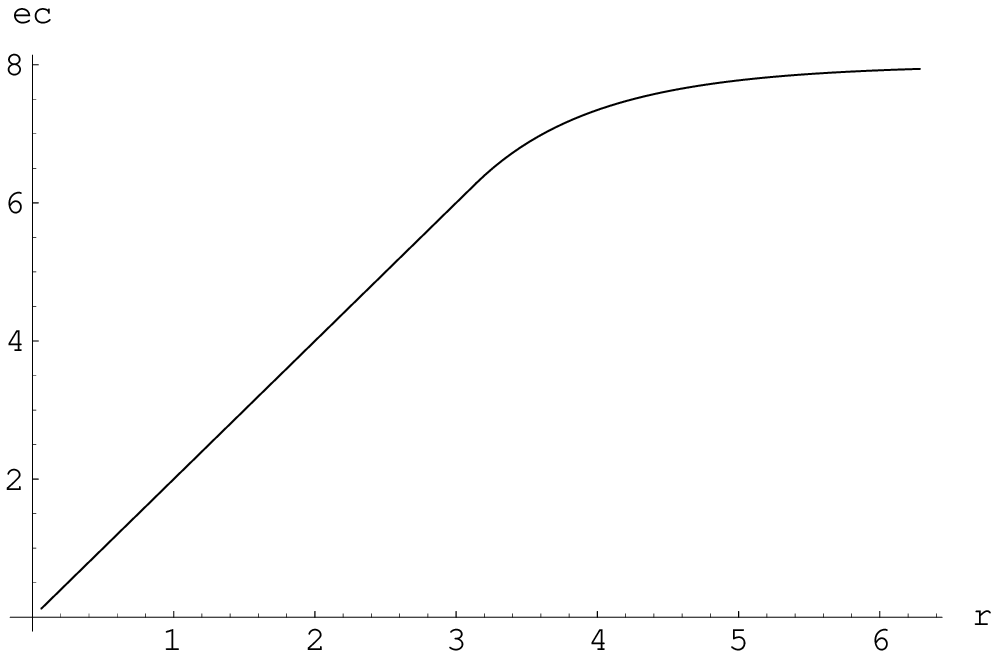,height=5cm,width=6.5cm} \caption{Classical
energy (\ref{SGclassencomplete})}\label{figclassenSG}
\end{center}
\end{figure}

We will now present a similar analysis for the solution
(\ref{SGsn}), which satisfies antiperiodic b.c.
\begin{equation}\label{antiperbcSG}
\phi(x+R)\,=\,-\,\phi(x)+\frac{2\pi}{\beta}\;,
\end{equation}
if it is defined on a cylinder of circumference
\begin{equation}\label{sizeSG}
R=\frac{1}{m}\,2 \textbf{K}\left(k^{2}\right)\;.
\end{equation}
Similarly to the kink (\ref{phi4kink}) studied in the $\phi^4$
case, the solution (\ref{SGsn}) tends to the standard infinite--volume soliton
(\ref{SGkinkinfvol}) for $A\rightarrow 0$, when $R$ goes to
infinity. In the other limit $A\rightarrow -2$, which corresponds
to $m R \rightarrow \pi$, (\ref{SGsn}) goes to the constant
solution
\begin{equation}\label{SGkinkUV}
\phi_{cl}(x)\equiv\frac{\pi}{\beta}\;,
\end{equation}
which identically satisfies the antiperiodic b.c.
(\ref{antiperbcSG}) and can be therefore used as the background in
the interval $0<mR<\pi$. Therefore, the classical energy
associated to this kink background is
\begin{equation}\label{SGclassencomplete}
{\cal
E}_{cl}(R)=\begin{cases}\frac{8m}{\beta^2}\left[\textbf{E}(k)-\frac{1}{2}(1-k^2)\textbf{K}(k)\right]
&\quad\text{for} \quad
mR>\pi\\
\frac{2m}{\beta^2}\, mR &\quad\text{for} \quad
mR<\pi\end{cases}\;,
\end{equation}
and it is plotted in Fig.\,\ref{figclassenSG}.

\vspace{0.5cm}

\psfrag{ec}{$\frac{{\cal E}_{cl}}{m/\beta^2}$}\psfrag{r}{$r$}

\begin{figure}[h]
\hspace{3cm} \psfig{figure=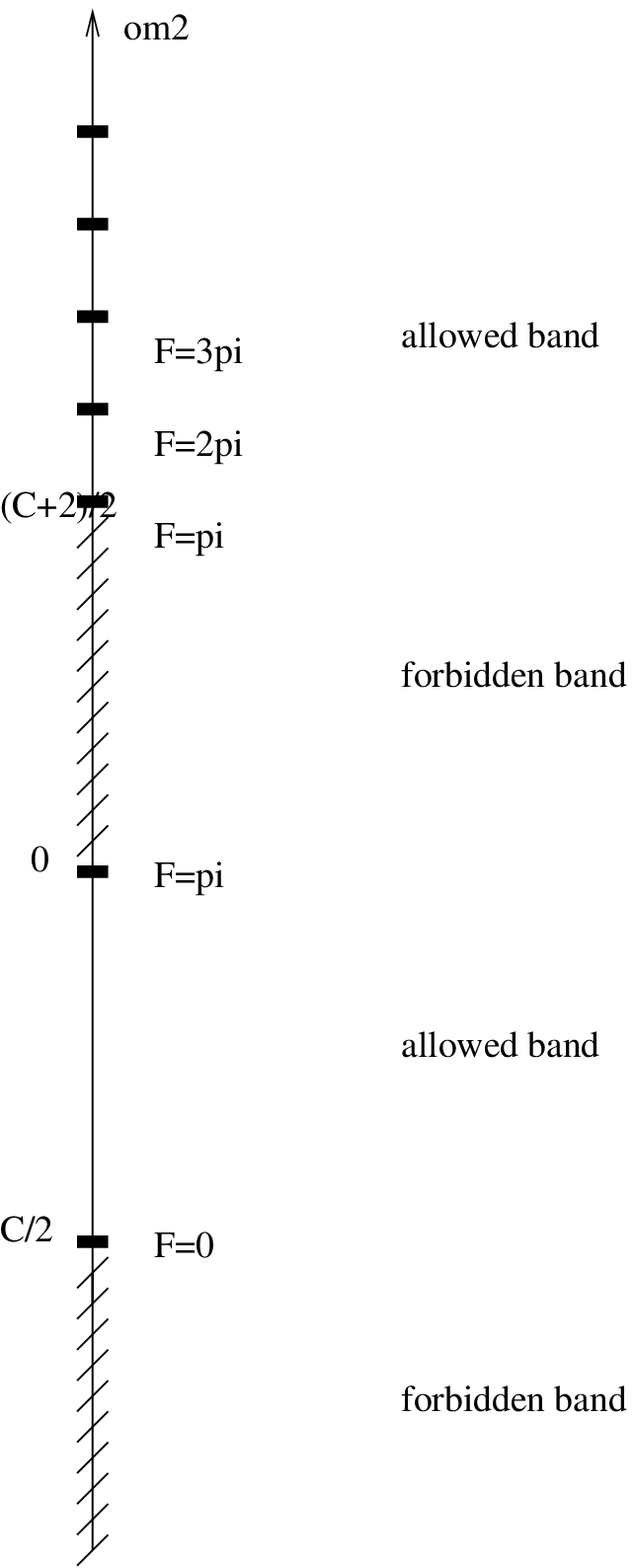,height=7cm,width=10cm}
\vspace{0.1cm} \caption{Spectrum of eq.\,(\ref{SGsemicl})}
\label{spectrumSG}
\end{figure}

The stability equation associated to (\ref{SGsn}) takes the form
\begin{equation}\label{SGsemicl}
\left\{\frac{d^{2}}{d \bar{x}^{2}}+\bar{\omega}^{2}+1-2
k^{2}\,\textrm{sn}^{2}\bar{x} \right\}\eta(\bar{x})=0\;,
\end{equation}
where
\begin{equation}
\bar{x}=m x\,,\quad \bar{\omega}=\frac{\omega}{m}\,.
\end{equation}
This can be cast in the Lam\'e form with $N=1$ (for the details,
see Appendix \ref{lame}), which has the band structure shown in
Fig.\,\ref{spectrumSG}. Imposing then the antiperiodic boundary
conditions (i.e. selecting the values of $\bar{\omega}^{2}$ for
which the Floquet exponent is an odd multiple of $\pi$), we obtain
the simple eigenvalues $\bar{\omega}_{0}^{2}=0$ and
\begin{equation}\label{omegadSG}
\bar{\omega}_{1}^{2}=k\;,
\end{equation}
and the infinite series of double eigenvalues
\begin{equation}\label{omeganSG}
\bar{\omega}_{n}^{2}\equiv \frac{2k^{2}-1}{3}-{\cal P}(i y_{n})\;
\end{equation}
in the band $\bar{\omega}^{2}>k^{2}$, with $y_{n}$ defined by
\begin{equation}\label{ynSG}
F=2\textbf{K}\,i\,\zeta(i y_{n})+2 y_{n}\,\zeta(\textbf{K})=(2
n-1)\pi\;, \qquad\quad n=2,3...
\end{equation}

\psfrag{om2}{$\bar{\omega}^{2}$}
\psfrag{(C+2)/2}{$\hspace{0.4cm}k^{2}$} \psfrag{0}{$0$}
\psfrag{C/2}{$\hspace{-0.4cm}k^{2}-1$} \psfrag{F=0}{$F=0$}
\psfrag{F=pi}{$F=\pi$}\psfrag{F=2pi}{$F=2\pi$}\psfrag{F=3pi}{$F=3\pi$}
\psfrag{allowed band}{allowed band}\psfrag{forbidden
band}{forbidden band}

\begin{figure}[ht]
\begin{center}
\psfig{figure=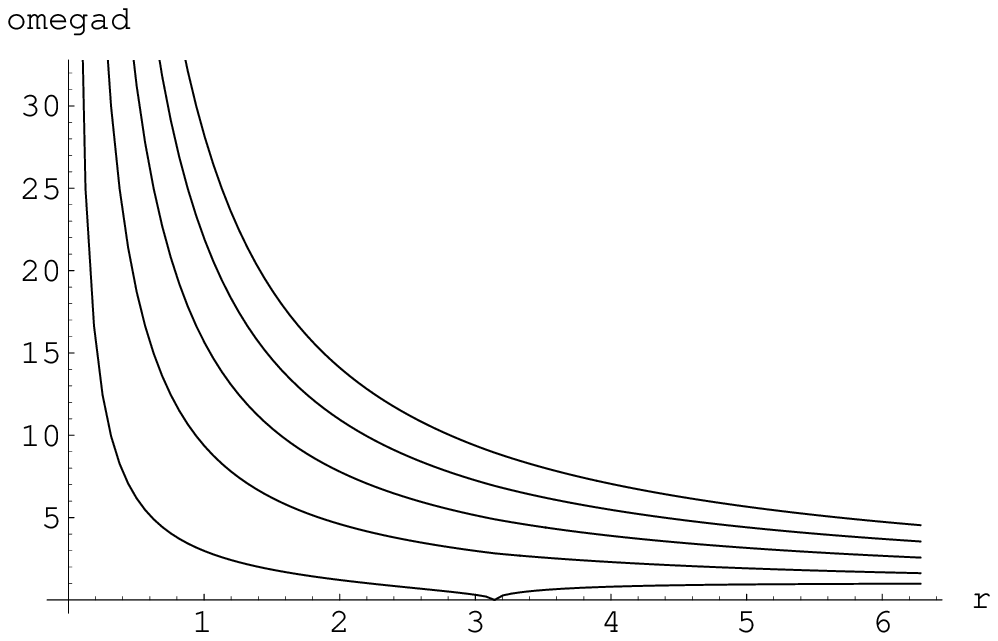,height=6cm,width=9cm} \caption{The
first few levels defined in (\ref{omegadcompleteSG}) and
(\ref{omegancompleteSG})}\label{figomegaiSG}
\end{center}
\end{figure}

It is easy to see that in the IR limit ($A\rightarrow 0 $) this
spectrum goes to the one related to the standard background
(\ref{SGkinkinfvol}). In order to complete the spectrum, also at
values $mR<\pi$, we have to glue the frequencies (\ref{omegadSG})
and (\ref{omeganSG}) with the ones obtained by quantizing the
constant solution (\ref{SGkinkUV}). We therefore obtain
\begin{equation}\label{omegadcompleteSG}
\bar{\omega}_{1}^{2}=\begin{cases}k&\qquad\text{for} \qquad mR>\pi\\
-1+\frac{\pi^2}{m^2 R^2}&\qquad\text{for} \qquad
mR<\pi\end{cases}\;,
\end{equation}
and
\begin{equation}\label{omegancompleteSG}
\bar{\omega}_{n}^{2}= \begin{cases}\frac{2k^{2}-1}{3}-{\cal P}(i
y_{n})
&\qquad\text{for} \qquad mR>\pi\\
-1+(2n-1)^2\frac{\pi^2}{m^2 R^2}&\qquad\text{for} \qquad
mR<\pi\end{cases}\;,
\end{equation}
which are plotted in Fig.\,\ref{figomegaiSG}.

\vspace{0.5cm}

\psfrag{omegad}{$\omega_{i}/m$}\psfrag{r}{$r$}

The study of the corresponding scaling functions can be performed
along the same lines illustrated for the broken $\phi^4$ theory.
One easily obtains the UV limit of the ground state energy in the
form
\begin{equation}
E_0(R) \,\simeq\,\frac{2\pi}{R} \left(h + \bar{h} -
\frac{c}{12}\right) + {\cal B} \, R + \cdots\; ,
\end{equation}
with $h + \bar{h}=1/8$ and
$$
{\cal B}\,=\,m^2\left(\frac{2}{\beta^2}\,-\,\frac{\log
2}{2\pi}\right)\;.
$$

Therefore, we have seen explicitly how the two types of kink
(\ref{SGam}) and (\ref{SGsn}), although they have the same IR
limit, display different energy levels in finite volume, and in
particular different UV limits, describing both twisted and
untwisted sectors of $c=1$ CFT.

\section{Open problems and discussion}\label{sectphiN}
\setcounter{equation}{0}

In this paper we have applied the semiclassical method to derive
analytic expressions for the energy levels of the broken $\phi^4$
theory on a cylinder with antiperiodic b.c.. Although this
analysis is technically similar to the one performed in
\cite{SGscaling} for the Sine--Gordon model in the one--kink
sector, various conceptual differences have emerged.

The derivation of analytic expressions for the finite--volume
semiclassical energy levels in the $\phi^4$ model is based on two
important ingredients: the explicit form of the kink solution
(\ref{phi4kink}) and the eigenvalues (\ref{omegad},\,\ref{omegan})
of the $N=2$ Lam\'e equation. Therefore its extension to $\phi^6$
and higher order $p\geq 5$ LG potentials (\ref{LGpot}) requires
the knowledge of the corresponding finite--volume kinks as well as
certain properties of the solutions of their stability equations
(\ref{stability}). Consider a family of symmetric (or \lq\lq
hyperelliptic") LG potentials
\begin{eqnarray}\nonumber
V_p^{\text{odd}}&=&\frac{1}{2}\,\prod\limits_{k=1}^{\frac{p-1}{2}}\,\left(\phi^2-a_k^2\right)^2
\qquad \quad \text{for}\quad p=3,5,...\\
V_p^{\text{even}}&=&\frac{1}{2}\,\phi^2\,\prod\limits_{k=1}^{\frac{p-2}{2}}\,\left(\phi^2-a_k^2\right)^2
\qquad \text{for}\quad p=4,6,... \label{LGpot2}
\end{eqnarray}
Their static kink solutions, i.e. the solutions of the first order
equation
$$
\frac{1}{2}\,\left(\frac{d\phi_{\text{cl}}}{dx}\right)^2\,=\,V_p(\phi_{\text{cl}})+A\,=
\,\frac{1}{2}\prod\limits_{l=1}^{p-1}\,\left(\phi_{\text{cl}}^2-b_l\right)\;,
$$
where $A=-V(\phi_0)$, $b_l=b_l(a_k)$ and $b_1=\phi_0^2$, are given
for both odd and even $p$ by the inverse of the following
hyperelliptic integrals:
\begin{equation}\label{implicitintegral}
\pm
2\,x\,=\,\int\limits_{\phi_0^2}^{\phi_{\text{cl}}^2(x)}\;\frac{dz}{\sqrt{\,z\,\prod\limits_{l=1}^{p-1}\,
\left(z-b_l\right)}}\;\;.
\end{equation}
In the case $p=4$ (i.e. for the $\phi^6$ model) the integral in
(\ref{implicitintegral}) is of elliptic type and the corresponding
finite--volume kink has the explicit form
\begin{equation}\label{phi6kink}
\phi_{\text{cl}}^{(p=4)}(x)\,=\,\frac{
\sqrt{b_1}}{\sqrt{1-\left(1-\frac{b_1}{b_2}\right)\text{sn}^2\left(\sqrt{b_2(b_3-b_1)}\;x,\,k\right)}}\;\;,
\end{equation}
where
$$
k^2\,=\,\left(\frac{b_3}{b_2}\right)\,\frac{b_2-b_1}{b_3-b_1}\;\;,
$$
$$
\begin{cases}\;b_2=\frac{1}{2}\left(2 a_1^2-b_1-\sqrt{b_1(4 a_1^2-3b_1)}\right)\\
\;b_3=\frac{1}{2}\left(2 a_1^2-b_1+\sqrt{b_1(4 a_1-3b_1)}\right)
\end{cases}\;\;.
$$
This background satisfies the boundary conditions
$$
\phi_{\text{cl}}(R)=\sqrt{b_1}+\sqrt{b_2}-\phi_{\text{cl}}(0)\;,
$$
provided we identify the size of the system as
$$
R=\frac{1}{\sqrt{b_2(b_3-\phi_0^2)}}\;\textbf{K}\;.
$$
Although for $p>4$ the kink solutions are not given in an explicit
form, one can easily derive their stability equation through the
change of variable $z=\phi_{\text{cl}}^2(x)$:
\begin{equation}\label{stabilityphiN}
\frac{d^2\eta(z)}{dz^2}+\frac{1}{2}\left(\frac{1}{z}+\sum\limits_{l=1}^{p-1}\frac{1}{z-b_l}\right)\frac{d\eta(z)}{dz}-
\frac{V_p''(z)-\omega^2}{2z\prod\limits_{l=1}^{p-1}(z-b_l)}\;\eta(z)=0\;\;,
\end{equation}
with the antiperiodic b.c. expressed as
\begin{equation}\label{antiperbcphiN}
\eta(z(R))=-\eta(z(0))\;\;,
\end{equation}
where $R$ is the smallest real period of the hyperelliptic
integral (\ref{implicitintegral}). The above second order ODE's
with $p+1$ regular singular points (at $z=0,\,b_l,\,\infty$)
represents a generalization \cite{ince} of the Lam\'e equation in
the so--called algebraic form\footnote{The same equation is
expressed in the alternative Weierstrass form in (\ref{lameeq}).}
\begin{equation}\label{lamealg}
\frac{d^2\eta(z)}{dz^2}+\frac{1}{2}\left(\frac{1}{z}+\frac{1}{z-1}+\frac{1}{z-a}\right)\frac{d\eta(z)}{dz}-
\frac{N(N+1)\,z-\lambda}{2z\,(z-1)(z-a)}\;\eta(z)=0\;\;,
\end{equation}
which coincides with (\ref{stabilityphiN}) for $N=2$, $p=3$ and
$V_3''(z)=6z-2 a_1^2$, i.e. for the $\phi^4$ potential analyzed in
Sect.\,\ref{sectphi4}.

Hence the derivation of the semiclassical scaling functions of the
generic $p\geq 4$ LG models (\ref{LGpot2}) defined on the cylinder
reduces to the problem of construction of the solutions and
eigenvalues of the generalized Lam\'e equation
(\ref{stabilityphiN}) for antiperiodic b.c. (\ref{antiperbcphiN}).
For $p\geq 4$ this is an interesting open problem, whose analytical
or numerical solutions will provide the necessary ingredients
for calculations of the corresponding energy levels.

Finally, it is worth mentioning few more research directions that
arise as natural developments of the analysis carried out here.
One of them consists of the determination of the energy levels in
the presence of different boundary conditions. Equally interesting
is to extend our computations to higher loop orders: although the
one--loop quantization around a kink background is more powerful
than standard perturbative techniques, we have seen however that
it is not yet accurate enough to identify the Ising critical point
in the phase diagram of the $\phi^4$ theory. The last point we
would like to mention is the study of symmetry restoration in
finite volume for antiperiodic boundary conditions. This
phenomenon is well understood in the vacuum sector (i.e. for
periodic b.c. \cite{brezin,zinnjustin}) but it is still an open
problem in the kink sector, and it may be fruitfully investigated
within the semiclassical approach.

\vspace{1cm}

\begin{flushleft}\large
\textbf{Acknowledgements}
\end{flushleft}
This work is done under the European Commission TMR programme
HPRN-CT-2002-00325 (EUCLID). The work of VR is supported by EPSRC,
under the grant GR/R83712/01. One of us (GM) would like to thanks
LPTHE and the ENS in Paris for their warm hospitality during the
period in which this work was completed, and CNRS for partial
financial support.

\vspace{1cm}

\begin{appendix}

\section{Elliptic integrals and Jacobi's elliptic
functions}\label{appell}\setcounter{equation}{0}

In this appendix we collect the definitions and basic properties
of the elliptic integrals and functions used in the text.
Exhaustive details can be found in \cite{GRA}.

The complete elliptic integrals of the first and second kind,
respectively, are defined as
\begin{equation}\label{ellint}
\textbf{K}(k^{2})\,=\,\int\limits_{0}^{\pi/2}\frac{d\alpha}
{\sqrt{1-k^{2}\sin^{2}\alpha}}\;,\qquad
\textbf{E}(k^{2})\,=\,\int\limits_{0}^{\pi/2}d\alpha
\sqrt{1-k^{2}\sin^{2}\alpha}\;.
\end{equation}
The parameter $k$, called elliptic modulus, has to be bounded by
$k^{2} < 1$. It turns out that the elliptic integrals are nothing
but specific hypergeometric functions, which can be easily
expanded for small $k$:
\begin{eqnarray*}
\textbf{K}(k^{2})&=&\frac{\pi}{2}\;F\left(\frac{1}{2},\frac{1}{2},1;k^{2}\right)=
\frac{\pi}{2}\,\left\{1+\frac{1}{4}\,k^{2}+\frac{9}{64}\,k^{4} +
\ldots + \left[\frac{(2n-1)!!}{2^{n}n!}\right]^{2}k^{2n} + \ldots
\right\}\;,\\
\textbf{E}(k^{2})&=&\frac{\pi}{2}\;F\left(-\frac{1}{2},\frac{1}{2},1;k^{2}\right)=
\frac{\pi}{2}\,\left\{1-\frac{1}{4}\,k^{2}-\frac{3}{64}\,k^{4}+
\ldots -
\left[\frac{(2n-1)!!}{2^{n}n!}\right]^{2}\frac{k^{2n}}{2n-1} +
\ldots \right\}\;.
\end{eqnarray*}
Furthermore, for $k^{2}\to 1$, they admit the following expansion
in the so--called complementary modulus $k' = \sqrt{1-k^{2}}$:
\begin{eqnarray*}
\textbf{K}(k^{2})&=&
\log\frac{4}{k'}+\left(\log\frac{4}{k'}-1\right)\frac{k'^{2}}{4} +
\ldots\;,\\\nonumber \textbf{E}(k^{2})&=&1+\left(\log\frac{4}{k'}
- \frac{1}{2}\right)\frac{k'^{2}}{2} + \ldots \;.
\end{eqnarray*}
Note that the complementary elliptic integral of the first kind is
defined as
$$
\textbf{K}'(k^{2}) \,=\, \textbf{K}(k'^{2})\;.
$$

The function $\text{am}(u,k^{2})$, depending on the parameter $k$,
and called Jacobi's elliptic amplitude, is defined through the
first order differential equation
\begin{equation}\label{ellam}
\left(\frac{d\,\text{am}(u)}{du}\right)^{2}\,=\, 1 - k^{2}
\sin^{2} \left[\text{am}(u)\right]\;,
\end{equation}
and it is doubly quasi--periodic in the variable $u$:
$$
\text{am}\left(u+2n\textbf{K}+2im\textbf{K}'\right) \,= \,
n\pi+\text{am}(u)\;.
$$
The Jacobi's elliptic function $\text{sn}(u,k^{2})$, defined
through the equation
\begin{equation}\label{ellsn}
\left(\frac{d\,\text{sn}u}{du}\right)^{2} \,=\,
\left(1-\text{sn}^{2}u\right)\left(1-k^{2}\text{sn}^{2}u\right)\;,
\end{equation}
is related to the amplitude by
$\text{sn}\,u=\sin\left(\text{am}\,u\right)$, and it is doubly
periodic:
$$
\text{sn}\left(u+4
n\textbf{K}+2im\textbf{K}'\right)\,=\,\text{sn}(u)\;.
$$

\section{Lam\'e equation}\label{lame}
\setcounter{equation}{0}

The second order differential equation
\begin{equation}\label{lameeq}
\left\{\frac{d^{2}}{d u^{2}}-E-N(N+1){\cal P}(u)\right\}f(u) \,=\,
0\;,
\end{equation}
where $E$ is a real quantity, $N$ is a positive integer and ${\cal
P}(u)$ denotes the Weierstrass function, is known under the name
of $N$-th Lam\'e equation. The function ${\cal P}(u)$ is a doubly
periodic solution of the first order equation (see \cite{GRA})
\begin{equation}\label{defP}
\left(\frac{d{\cal P}}{du}\right)^{2}\,=\,4\left({\cal P} -
e_{1}\right)\,\left({\cal P} - e_{2}\right)\,\left({\cal P} -
e_{3}\right)\;,
\end{equation}
whose characteristic roots $e_{1},e_{2},e_{3}$ uniquely determine
the half--periods $\omega$ and $\omega'$, defined by
$$
{\cal P}\left(u+2n\omega+2 m\omega'\right)\,=\,{\cal P}(u)\;.
$$

The stability equation (\ref{phi4semicl}), related to the broken
$\phi^4$ theory, can be identified with eq. (\ref{lameeq}) for
$N=2$, $u=\frac{\bar{\phi}_0}{\sqrt{2}}\,\bar{x}+i\textbf{K}'$ and
$E=(1+k^2)(1-\bar{\omega}^{2})$; also the stability stability
equation (\ref{SGsemicl}), encountered in the analysis of the
Sine--Gordon model, can be identified with eq. (\ref{lameeq}), in
this case with $N=1$, $u=\bar{x}+i\textbf{K}'$ and $E=\frac{2
k^{2}-1}{3}-\bar{\omega}^{2}$. Both these identifications hold in
virtue of the relation between ${\cal P}(u)$ and the Jacobi
elliptic function $\text{sn}(u,k)$ (see formulas 8.151 and 8.169
of \cite{GRA}):
\begin{equation}\label{Psn}
k^{2}\text{sn}^{2}(\bar{x},k) \,=\,{\cal P}(\bar{x}+i\textbf{K}')
+ \frac{k^{2}+1}{3}\;.
\end{equation}
Relation (\ref{Psn}) is valid if the characteristic roots of
${\cal P}(u)$ are expressed in terms of $k^{2}$ as
\begin{equation}\label{roots}
e_{1}\,=\,\frac{2-k^{2}}{3}\;, \qquad
e_{2}\,=\,\frac{2k^{2}-1}{3}\;,\qquad e_{3}
\,=\,-\frac{1+k^{2}}{3}\;,
\end{equation}
and, as a consequence, the real and imaginary half periods of
${\cal P}(u)$ are given by the elliptic integrals of the first
kind
\begin{equation}\label{halfper}
\omega \,=\,\textbf{K}(k)\;,\qquad \omega' \,=\, i \textbf{K}'(k)
\;.
\end{equation}
All the properties of Weierstrass functions that we will use in
the following are specified to the case when this identification
holds.

We will now present the solutions of the Lam\'e equation for $N=1$
and $N=2$, which have been derived in \cite{tak,smirnov} together
with more complicated cases.

In the case $N=1$ the two linearly independent solutions of
(\ref{lameeq}) are given by
\begin{equation}
f_{\pm a}(u)\,=\,\frac{\sigma(u\pm a)}{\sigma(u)}\;e^{\mp\,u
\,\zeta(a)}\;,
\end{equation}
where $a$ is an auxiliary parameter defined through ${\cal
P}(a)=E$, and $\sigma(u)$ and $\zeta(u)$ are other kinds of
Weierstrass functions:
\begin{equation}\label{defsigma}
\frac{d\,\zeta(u)}{du} \,=\, - {\cal P}(u)\;,\qquad
\frac{d\,\log\sigma(u)}{du} \,=\, \zeta(u)\;,
\end{equation}
with the properties
\begin{eqnarray}\nonumber
&&\zeta(u+2\textbf{K}) \,=\,\zeta(u) + 2
\zeta(\textbf{K})\;,\\\label{zsprop}
&&\sigma(u+2\textbf{K})\,=\,-\,e^{2(u+\textbf{K})
\zeta(\textbf{K})} \sigma(u)\;.
\end{eqnarray}
As a consequence of eq. (\ref{zsprop}) one obtains the Floquet
exponent of $f_{\pm a}(u)$, defined as
\begin{equation}
f(u+2 \textbf{K}) \,=\, f(u)e^{i F(a)}\;,
\end{equation}
in the form
\begin{equation}
F(\pm a) \,=\, \pm 2 i\left[\textbf{K}\,\zeta(a) -
a\,\zeta(\textbf{K})\right]\;.
\end{equation}
The spectrum in the variable $E$ of eq. (\ref{lameeq}) with $N=1$
is divided in allowed/forbidden bands depending on whether $F(a)$
is real or complex for the corresponding values of $a$. We have
that $E < e_{3}$ and $e_{2} < E < e_{1}$ correspond to allowed
bands, while $e_{3} < E < e_{2}$ and $E > e_{1}$ are forbidden
bands. Note that if we exploit the periodicity of ${\cal P}(a)$
and redefine $a\rightarrow a' = a + 2 n \omega+2 m \omega'$, this
only shifts $F$ to $F' = F+2 m \pi$.

\begin{figure}[h]
\hspace{1.5cm} \psfig{figure=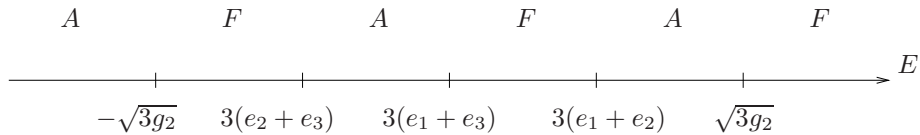,height=1.5cm,width=12cm}
\caption{Spectrum of eq.\,(\ref{lameeq}) with $N=2$, where
$e_1,\,e_2,\,e_3$ are the roots of ${\cal P}$ and
$g_2=2(e_1^2+e_2^2+e_3^2)$.} \label{figE}
\end{figure}

The solutions of the Lam\'e equation with $N=2$ are given by
\begin{equation}
f(u)=\frac{\sigma(u+ a)\,\sigma(u+ b)}{\sigma^{2}(u)}\;e^{- u
\,[\zeta(a)+\zeta(b)]}\;,
\end{equation}
where $a$ and $b$ are two auxiliary parameters satisfying the
constraints
\begin{equation}\label{constr}
\begin{cases}3{\cal P}(a)+3{\cal P}(b)=E\\
{\cal P}'(a)+{\cal P}'(b)=0\end{cases}\;,
\end{equation}
and $\sigma(u)$ and $\zeta(u)$ are defined in (\ref{defsigma}).
The Floquet exponent of $f(u)$ is now given by
\begin{equation}
F= 2
i\left\{\textbf{K}[\zeta(a)+\zeta(b)]-(a+b)\zeta(\textbf{K})\right\}\;.
\end{equation}
The spectrum in the variable $E$ of eq. (\ref{lameeq}) with $N=2$
is divided in allowed (A) and forbidden (F) bands depending on
whether $F$ is real or complex for the corresponding values of $a$
and $b$, as shown in Fig.\,\ref{figE}.

\vspace{1cm}

\psfrag{E}{\small$E$}\psfrag{e1}{\hspace{-0.7cm}\small$-\sqrt{3
g_2}$}\psfrag{e2}{\hspace{-1cm}\small$3(e_2+e_3)$}\psfrag{e3}{\hspace{-0.8cm}\small$3(e_1+e_3)$}
\psfrag{e4}{\hspace{-0.5cm}\small$3(e_1+e_2)$}
\psfrag{e5}{\hspace{-0.3cm}\small$\sqrt{3
g_2}$}\psfrag{A}{\small$A$}\psfrag{F}{\small$F$}

\vspace{0.5cm}

Finally, it is worth mentioning that the function $\zeta(u)$
admits a series representation \cite{hancock} that is very useful
for our purposes in the text:
\begin{equation}
\zeta(u)\,=\,\frac{\pi}{2\textbf{K}}\,\cot\left(\frac{\pi u}
{2\textbf{K}}\right)+\left(\frac{\textbf{E}}{\textbf{K}}
+\frac{k^{2}-2}{3}\right)\,u + \frac{2\pi}{\textbf{K}}
\sum\limits_{n=1}^{\infty} \frac{h^{2n}}{1-h^{2n}}\,\sin
\left(\frac{n \pi u}{\textbf{K}}\right)\;,
\end{equation}
where $h = e^{-\pi\textbf{K}'/\textbf{K}}$. The small-$k$
expansion of this expression gives
\begin{equation}
\label{expzeta} \hspace{-3.9cm}\zeta(u)\,=\,\left(\cot u +
\frac{u}{3}\right) \,+\,\frac{k^{2}}{12}\left(u - 3 \cot u + 3 u
\cot^{2}u\right)\,+
\end{equation}
\begin{equation*}
+\,\frac{k^{4}}{64}\left(-3 u + ( 4 u^{2} - 5 ) \cot u + u
\cot^{2}u + 4 u^{2} \cot^{3}u + \sin 2u \right) + \ldots
\end{equation*}
(note that $h\approx
\left(\frac{k}{4}\right)^{2}+O\left(k^{4}\right)$). A similar
expression takes place for ${\cal P}(u)$, by noting that ${\cal
P}(u) = - \frac{d\,\zeta(u)}{du}$.

\section{Ground state energy regularization at $r\approx\pi$}\label{appfinite}\setcounter{equation}{0}

We present in this appendix the evaluation of the ground state
energy (\ref{grstate}) for $r\lesssim\pi$ and $r\gtrsim\pi$,
comparing the two corresponding expressions at the point $r=\pi$.

In the case $r\lesssim\pi$, we obtain
\begin{equation}\label{grstateleft}
\frac{E_0}{m}(r)\,=\,A_-\,+\,\sqrt{2}\,\sqrt{1-\frac{r}{\pi}}\,+\,B_-\,\left(1-\frac{r}{\pi}\right)\,+\,...\;,
\end{equation}
where the coefficients $A_-$ and $B_-$ are defined as
\begin{eqnarray*}
A_-&=&\frac{m^2}{\lambda}\,\frac{\pi}{4}\,+\,\sum\limits_{n=1}^{\infty}
\sqrt{(2n-1)^2-1}\,+\,\frac{3}{2}\,\sum\limits_{n=1}^{\infty}\frac{1}{\sqrt{(2n)^2+2}}
\,-\,\sum\limits_{n=1}^{\infty}\sqrt{(2n)^2+2}\,-\,\frac{1}{4\sqrt{2}}\;\;,\\
B_-&=&-\frac{m^2}{\lambda}\,\frac{\pi}{4}\,+\,\sum\limits_{n=2}^{\infty}
\frac{(2n-1)^2}{\sqrt{(2n-1)^2-1}}\,-\,\frac{3}{2}\,\sum\limits_{n=1}^{\infty}\frac{(2n)^2}{[(2n)^2+2]^{3/2}}
\,-\,\sum\limits_{n=1}^{\infty}\frac{(2n)^2}{\sqrt{(2n)^2+2}}\;\;.
\end{eqnarray*}
Expanding in $\frac{1}{(2n-1)}$ and $\frac{1}{(2n)}$, we obtain
\begin{eqnarray}\nonumber
A_-&=&\frac{m^2}{\lambda}\,\frac{\pi}{4}\,+\,\sum\limits_{n=1}^{\infty}
(2n-1)\,-\,\sum\limits_{n=1}^{\infty}(2n)\,-\,\frac{1}{2}\sum\limits_{n=1}^{\infty}
\frac{1}{(2n-1)}\,+\,\frac{1}{2}\,\sum\limits_{n=1}^{\infty}\frac{1}{(2n)}
\,-\,\frac{1}{4\sqrt{2}}\,-\,C_-\;\;,\\
B_-&=&-\,\frac{m^2}{\lambda}\,\frac{\pi}{4}\,+\,\sum\limits_{n=1}^{\infty}(2n-1)\,-\,\sum\limits_{n=1}^{\infty}(2n)
\,+\,\frac{1}{2}\,\sum\limits_{n=1}^{\infty} \frac{1}{(2n-1)}
\,-\,\frac{1}{2}\,\sum\limits_{n=1}^{\infty}\frac{1}{(2n)}\,-\,1\,+\,D_-\;\;,
\label{AB-}
\end{eqnarray}
where $C_-$ and $D_-$ are finite constants given by
\begin{eqnarray*}
C_-&=&\sum\limits_{k=1}^{\infty}\,\frac{(-1)^{k}(2
k-1)!!}{(k+1)!\,2^{\,2k+1}}\,\left[\frac{\zeta(2k+1,1/2)}{2^{k+1}}\,-\,\frac{3k+1}{2}\,\zeta(2k+1)\right]\;\;,\\
D_-&=&\sum\limits_{k=1}^{\infty}\,\frac{(-1)^{k+1}(2
k-1)!!}{k!\,2^{\,2k+1}}\,\left[\frac{2k+3}{k+1}\,\frac{\zeta(2k+1,1/2)}{2^{k+1}}\,+\,\frac{(3
k+1)(2k+1)}{2(k+1)}\,\zeta(2k+1)\,-\,2^{k+1}\right]\;\;,
\end{eqnarray*}
with numerical values $C_-\simeq 0.018$, $D_-\simeq 0.39$, and the
functions in this expressions are defined as
\begin{equation*}
\begin{cases}\;n!\,=\,1\cdot 2\cdot \,\ldots\, \cdot n \\
\;(2n+1)!!\,=\,1\cdot 3\cdot \,\ldots\, \cdot
(2n+1)\end{cases},\qquad\qquad
\begin{cases}\;\zeta(p)\,=\,\sum\limits_{n=1}^{\infty}\frac{1}{k^{p}}\\
\;\zeta(p\,,\alpha)\,=\,\sum\limits_{n=0}^{\infty}\frac{1}{(k+\alpha)^{p}}\end{cases}\;.
\end{equation*}
The individually divergent series present in (\ref{AB-}) combine
to give a finite result, in virtue of the relations
\begin{eqnarray*}
&&\sum\limits_{n=0}^{\infty}
(2n+1)\,-\,\sum\limits_{n=1}^{\infty}(2n)\,=\,2
\left[\zeta(-1,1/2)-\zeta(-1)\right]\,=\,2\left[\frac{1}{24}
+\frac{1}{12}\right]\;,\\
&&
\sum\limits_{n=0}^{\infty}\frac{1}{(2n+1)}\,-\,\sum\limits_{n=1}^{\infty}\frac{1}{(2n)}\,=\,
\sum\limits_{k=1}^{\infty}\frac{(-1)^{k+1}}{k}\,=\,\log 2\;.
\end{eqnarray*}
Therefore, the final expressions for the coefficients $A_-$ and
$B_-$ are
\begin{eqnarray}\nonumber
A_-&=&\frac{m^2}{\lambda}\,\frac{\pi}{4}\,+\,\frac{1}{4}\,-\,\frac{1}{2}\log
2\,-\,\frac{1}{4\sqrt{2}}\,-\,C_-\;\;,\\
B_-&=&-\frac{m^2}{\lambda}\,\frac{\pi}{4}\,+\,\frac{1}{4}\,+\,\frac{1}{2}\log
2\,-\,1\,+\,D_-\;\;.\nonumber
\end{eqnarray}

\vspace{0.5cm}

The other case $mR\gtrsim\pi$ can be similarly treated, being more
complicated only from the technical point of view. In fact, it
requires to compare, in the limit $k\to 0$, the behavior of
classical energy and stability frequencies, defined in
(\ref{phi4classen}), (\ref{omegad}) and (\ref{omegan})
respectively, with the one of the scaling variable, defined in
(\ref{sizephi4}). The expansions of elliptic integrals and
Weierstrass functions, necessary for this purpose, can be found in
Appendices \ref{appell} and \ref{lame}. Since the scaling variable
has the small--$k$ behaviour
\begin{equation}\label{sizephi4smallk}
r\,=\,\pi\left[1+\frac{3}{4}\,k^2+...\right]\;,
\end{equation}
it is easy to see that
\begin{equation}
\frac{{\cal
E}_{\text{cl}}}{m}\,=\,\frac{m^2}{\lambda}\,\frac{\pi}{4}\,\left(1+\frac{3}{4}\,k^2\right)+...\,=\,
\frac{m^2}{\lambda}\,\frac{\pi}{4}+\frac{m^2}{4\lambda}\,(r-\pi)+...
\end{equation}
and
\begin{equation}
\frac{\omega_1}{m}\,=\,\sqrt{3}\,k+...\,=\,2\sqrt{\frac{r}{\pi}-1}+...\;.
\end{equation}
The frequencies (\ref{omegan}) have the most implicit expression
in term of $r$. Noting that in the highest band
$\;\bar{\omega}^{2}>1+\frac{2\sqrt{k^4-k^2+1}}{1+k^2}\;$ the
auxiliary paramaters $a$ and $b$ are related as $a=-b^*$, we can
conveniently parameterize $a_n$ and $b_n$ in (\ref{anbn}) as
\begin{equation}
\begin{cases}
a_n\,=\,-x_n+i y_n\\
b_n\,=\,x_n+i y_n\end{cases}
\end{equation}
Expanding equations (\ref{anbn}) for small $k$, we obtain
\begin{equation}
\begin{cases}
x_n\,=\,\frac{1}{2}\,\arcsin\left(\sqrt{\frac{3}{(2n+1)^2-1}}\;\right)\left[1+\frac{k^2}{4}+...\right]\vspace{0.3cm}\\
y_n\,=\,\frac{1}{2}\,\text{arcsinh}\left(3\sqrt{\frac{(2n+1)^2}{[(2n+1)^2-1][(2n+1)^2-4]}}\;\right)
\left[1+\frac{k^2}{4}+...\right]\end{cases}
\end{equation}
and therefore
\begin{equation}
\bar{\omega}_n^2\,=\,\left[(2n+1)^2-1\right]\left\{1-\frac{3}{2}\,k^2\;\frac{(2n+1)^2-2}{(2n+1)^2-1}+...\right\}\;.
\end{equation}
Comparing this with (\ref{sizephi4smallk}) we finally obtain
\begin{equation}\label{omeganexp}
\frac{\omega_n}{m}(r)\,=\,\sqrt{(2n+1)^2-1}\,-\,\frac{(2n+1)^2-2}{\sqrt{(2n+1)^2-1}}\left(\frac{r}{\pi}-1\right)\,+\,...\;.
\end{equation}
Therefore, the ground state energy has the behaviour
\begin{equation}\label{grstateright}
\frac{E_0}{m}(r)\,=\,A_+\,+\,\sqrt{\frac{r}{\pi}-1}\,+\,B_+\,\left(\frac{r}{\pi}-1\right)\,+\,...\;,
\end{equation}
where $A_+=A_-$ and $B_+=B_-$.

\end{appendix}

\end{document}